\documentclass[manuscript, nonacm]{acmart} 
\usepackage{graphicx} 
\usepackage{hyperref}

\usepackage{amsmath}
\usepackage{wrapfig}
\usepackage{listings}
\usepackage{xcolor}
\usepackage{multirow}
\usepackage{tcolorbox}
\usepackage{makecell}
\usepackage{tikz}
\lstdefinestyle{console}{
    basicstyle=\ttfamily\footnotesize,
    backgroundcolor=\color{gray!10},
    frame=single,
    breaklines=true,
    postbreak=\mbox{\textcolor{red}{$\hookrightarrow$}\space},
}
\newcommand{\rev}[1]{\textcolor{black}{#1}}
\newcommand{\revm}[1]{\textcolor{black}{#1}}
\usepackage{style/prompt}

\usepackage{style/codestyle}

\usepackage{multirow}
\usepackage{tabularx}
\usepackage{subcaption}
\usepackage{graphicx}
\usepackage{tcolorbox}
\usepackage{amsmath}
\usepackage{amsfonts}
\usepackage{fancyvrb}
\usepackage{xcolor}
\usepackage{tcolorbox}
\usepackage{wrapfig}

\newcounter{observation}
\renewcommand{\theobservation}{O\arabic{observation}}

\newtcolorbox{observation}[2][]{%
    colback=yellow!10!white,    
    colframe=black,             
    rounded corners,
    arc=5pt,
    boxrule=1pt,
    left=5pt,
    right=5pt,
    top=1pt,
    bottom=1pt,
    before skip=7pt,
    after skip=7pt,
    #1,                         
    before upper={%
        \refstepcounter{observation}
        \phantomsection
        \ifx&#2&\else
            \label{#2}
        \fi
        \noindent\textbf{Observation \theobservation: }%
    }
}

\newtcolorbox{takeaway}[1][]{
    colback=blue!10!white,    
    colframe=black,             
    rounded corners,            
    arc=5pt,                   
    boxrule=1pt,               
    left=5pt,                 
    right=5pt,                
    top=1pt,                   
    bottom=1pt, 
    before skip=7pt, 
    after skip=7pt,  
    label={#1}                         
}

\newcounter{k}
\newcounter{rec}
\newcommand{\recommendation}[1]{%
  \refstepcounter{rec}%
  \noindent\textbf{Recommendation (R\therec).} #1\par
}

\title{\textsc{BLAgent}: Agentic RAG for File-Level Bug Localization}

\titlenote{Accepted for publication in ACM Transactions on Software Engineering and Methodology (TOSEM).}

\author{Md Afif Al Mamun}
\email{afif.mamun@ucalgary.ca}
\affiliation{
    \institution{University of Calgary} 
    \department{Department of Electrical and Software Engineering}
    \city{Calgary}
    \country{Canada}
 } 
\author{Gias Uddin}
\email{guddin@yorku.ca}
\affiliation{
    \institution{York University}
    \department{Department of Electrical Engineering and Computer Science}
    \city{Toronto}
    \country{Canada}
 }

\begin{abstract}

Bug localization remains a key bottleneck for large language model (LLM)-based software maintenance, where accurately identifying faulty code is essential for debugging, root cause analysis, triage, and automated program repair (APR). File-level bug localization is especially critical in hierarchical localization and repair pipelines, where incorrect file selection can propagate to downstream stages such as function-level localization and patch generation. While Retrieval-Augmented Generation (RAG) offers a promising way to ground LLMs in repository context, existing RAG pipelines often rely on static retrieval and lack the reasoning needed to accurately identify faulty code. In this work, we present \textsc{\textbf{BLAgent}}, a novel agentic RAG framework for file-level bug localization that integrates three key ideas: (i) code structure-aware repository encoding with path-augmented AST-based chunking, (ii) dual-perspective query transformation that captures both structural and behavioral signals from bug reports, and (iii) two-phase agentic reranking that combines symbolic inspection with evidence-grounded reasoning. Unlike prior graph-based or multi-hop agentic approaches, \textsc{BLAgent} adopts a bounded reasoning strategy that limits LLM-based inspection and reranking to a compact, retrieval-filtered set of candidate files, avoiding open-ended repository traversal. This design balances localization accuracy with computational cost. On SWE-bench-Lite, \textsc{BLAgent} attains over 78\% Top-1 accuracy with open-source models and over 86\% with a closed-source model, while being over 18$\times$ cheaper than the strongest baseline using the same model. When integrated into an APR framework, \textsc{BLAgent} improves end-to-end repair success by up to 25\%.

\end{abstract}

\setcopyright{none}
\renewcommand\footnotetextcopyrightpermission[1]{}

\begin{document}



\maketitle

\section{Introduction}
\rev{Bugs are inevitable in software systems. When a bug is logged, various tasks take place, such as triaging, debugging, root cause analysis, and finally fixing. Across all these tasks, bug localization often remains the first step \cite{wong2016survey, wong2023software, bohme2017bug}. Bug localization (BL) is the process of identifying faulty code regions within a software repository. In large and evolving repositories, determining which files are most likely responsible for a reported issue can require substantial effort \cite{stripe2018developer, bohme2017bug}. Therefore, effective localization plays a broader role in software engineering by reducing the search space for both human and automated maintenance workflows. }

Traditional localization approaches often operate at multiple granularities, including file-level, function-level, and statement-level localization \cite{chang2025bridging, jiang2025cosil}. However, recent empirical evidence reveals that file-level localization is the most critical component in hierarchical bug localization pipelines \cite{chang2025bridging}. The study demonstrated that removing file-level localization from a multi-granularity localization framework caused a catastrophic 94\% drop in Top-5 accuracy and a 96\% reduction in Mean Average Precision (MAP) at the statement level. This finding suggests that without accurate file-level localization, even sophisticated localization techniques may fail to identify buggy code. 

Take SWE-bench \cite{jimenez2024swebench} as an example, a widely used Automated Program Repair (APR) benchmark for large language model (LLM)-based coding assistants. SWE-bench averages over 11,000 functions and 168,000 statements—direct statement-level localization is computationally infeasible, particularly for LLM-based systems \cite{chang2025bridging}. File-level localization, by contrast, reduces the search space by several orders of magnitude, enabling efficient downstream localization and repair. Consequently, improving file-level accuracy offers the most impactful path toward advancing end-to-end APR performance. Recent advances in LLMs have significantly improved APR capabilities on real-world bugs \cite{wang2024openhands, yang2024swe, xia2025agentless}. However, these systems remain fundamentally constrained by localization quality, as inaccurate file selection directly limits repair effectiveness \cite{chang2025bridging, jiang2025cosil}. \rev{While bug localization also supports developer-centric activities such as manual debugging and root cause analysis, evaluating its effectiveness in such settings often requires subjective or task-specific studies. We therefore adopt APR as a representative downstream evaluation setting, where localization quality can be assessed objectively through patch correctness and issue resolution.}


\begin{figure}[htbp]
    \centering
    \includegraphics[width=0.6\linewidth]{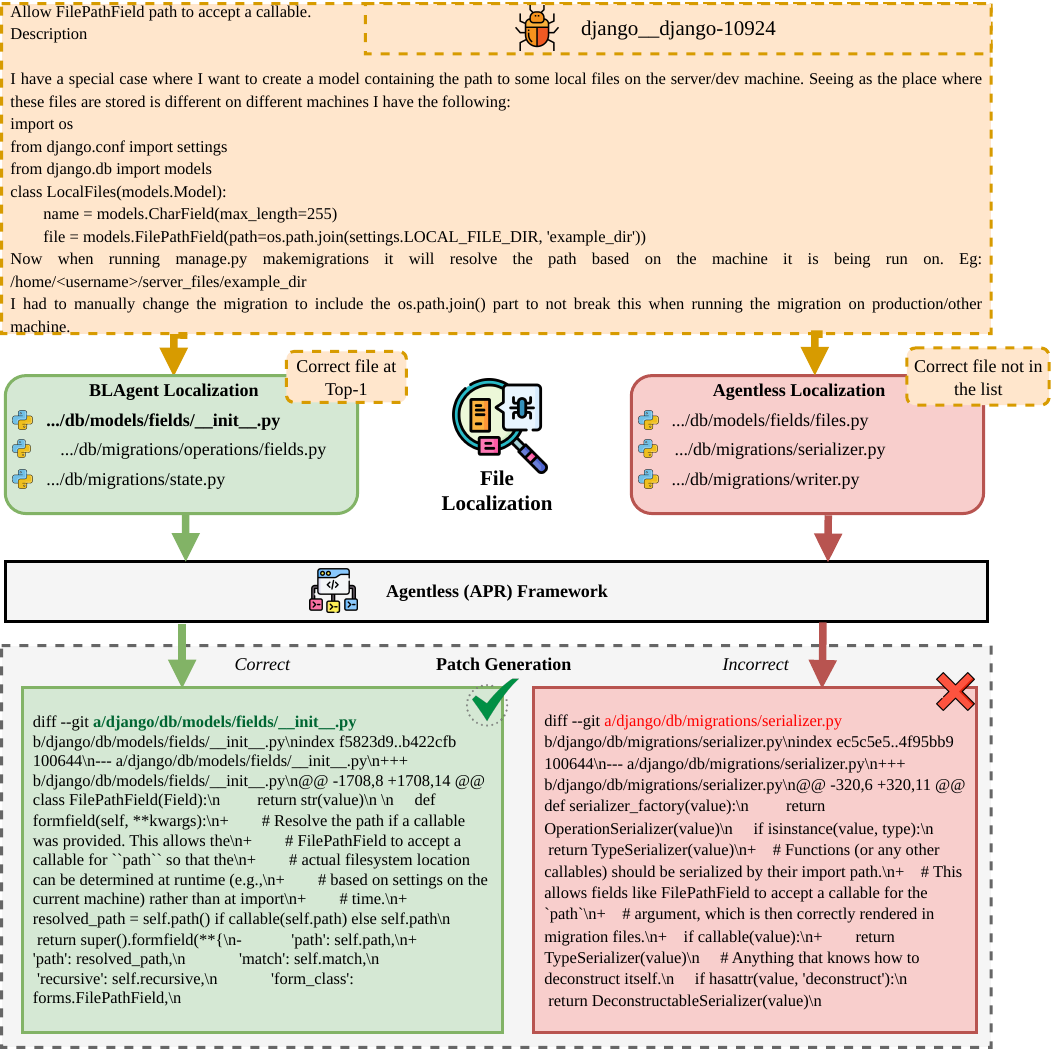}
    \caption{An example demonstrating how incorrect file localization may lead to incorrect patch generation.}
    \label{fig:localization-example}
\end{figure}

Consider the motivating example illustrated in Figure \ref{fig:localization-example}, which demonstrates how incorrect file localization directly leads to incorrect patch generation in an example (\texttt{django-10924}) from the Django project in the SWE-bench-Lite dataset. The bug report describes the need to allow \texttt{FilePathField} to accept a callable for its path parameter, thereby enabling runtime resolution of file system locations. An existing APR system (Agentless \cite{xia2025agentless}) incorrectly localizes the fault to files such as \texttt{serializer.py}. This misdirection causes the system to generate a patch modifying the migration serializer to handle callable values but misses the core issue persisting in a different file. In contrast, when the correct file \texttt{\_\_init\_\_.py} is identified through more precise localization, the generated patch by the same APR resolves the callable path within the \texttt{FilePathField.formfield()} method, addressing the root cause of the issue. 

This example shows that even sophisticated repair systems fail when bug localization provides incorrect context, highlighting the critical need for improved localization strategies that can bridge the semantic gap between natural language bug reports and code-level implementations. While providing more candidate files might seem to mitigate localization errors, this approach may face critical constraints. LLMs are constrained by limited context windows, and processing more files individually incurs substantial computational overhead. Moreover, LLMs are also sensitive to the order of the inputs, where their ability to extract relevant information deteriorates when details appear in the middle or later of lengthy inputs \cite{liu2023lost, lu2021fantastically}. For such reasons, APRs typically restrict consideration to only the top-\textit{k} (e.g., Top-3) files to balance context length and computational feasibility~\cite{joshi2023repair, xia2025agentless}. 

Retrieval-Augmented Generation (RAG) \cite{lewis2020retrieval} offers a promising direction for addressing these localization challenges. RAG enhances LLMs by grounding their predictions in external knowledge retrieved from code repositories, thereby improving factuality and reducing hallucinations. Recent work has explored RAG for code-related tasks such as code completion and bug localization, demonstrating that retrieval-based approaches can effectively bridge the vocabulary gap between natural language bug descriptions and technical code artifacts \cite{jimenez2024swebench, zhang2025cast, zhao2025recode, tao2025retrieval}. However, conventional RAG strategies for code often rely on holistic code-text embeddings and naive text-based chunking, which may fail to capture the structural intricacies of code and result in suboptimal retrieval quality \cite{zhang2025cast}.

Furthermore, purely embedding-based retrieval approaches, while efficient, lack the reasoning capabilities necessary to assess the functional relevance and appropriateness of retrieved code. Recent advances in LLM-based agents--where LLMs autonomously plan, reason, and act using external tools—present an opportunity to enhance RAG systems with explicit code-level reasoning \cite{huang2024understanding, guo2024large}. Agentic approaches have shown promise in various software engineering tasks by combining iterative reasoning with tool usage, enabling more informed decision-making compared to static retrieval methods \cite{yang2024swe, wang2024openhands, bouzenia2024repairagent, chen2025locagent}. Recent agentic localization methods often employ graph-based repository traversal, which can incur prohibitive costs \cite{chen2025locagent, jiang2025cosil, wang2024openhands} or require extensive LLM-finetuning \cite{chang2025bridging, chen2025locagent, jiang2025cosil}.

\rev{Motivated by these limitations, we propose \textsc{BLAgent} (\underline{B}ug \underline{L}ocalization using \underline{Agent}ic RAG), a novel agentic RAG framework for file-level bug localization that addresses the key limitations of typical RAG pipelines. First, \textsc{BLAgent} encodes a code repository using path-augmented, abstract syntax tree (AST)-aware chunking where source files are split at syntactic boundaries (e.g., functions, classes, logical units) via AST analysis, and each chunk is prepended with its relative file path. This preserves semantic integrity within chunks while creating a direct alignment channel between path-like signals in bug reports (e.g., tracebacks, module references) and the code they describe. Second, it applies dual-perspective query transformation, decomposing each bug report into a structural query that highlights code entities such as function names, modules, and identifiers, and a behavioral query that captures runtime symptoms and expected vs. observed behavior. Issuing both queries independently for retrieval and merging their candidate sets ensures that the retrieval pool covers faults described by lexical reference and by observable behavior. Third, \textsc{BLAgent} performs two-phase agentic reranking: a ReAct-based reasoning agent iteratively inspects structural skeletons of candidate files and assigns relevance scores, followed by a single-shot evidence-anchored reranking step that expands only retriever-highlighted code regions within top-ranked files for a final, implementation-level ranking decision. Unlike graph-based agentic approaches, \textsc{BLAgent} bounds its reasoning to a compact, retrieval-filtered candidate set, avoiding the costly unbounded traversal of existing approaches. We evaluate \textsc{BLAgent} through two research questions:}

\noindent\textbf{RQ1. How effective is \textsc{BLAgent} over baselines for bug localization?}

\textsc{BLAgent} significantly outperforms both conventional RAG pipelines and state-of-the-art approaches — including complex, graph-based methods — achieving an MRR of 0.900 and a Top-1 accuracy of 86.7\% with a closed-source model, surpassing LocAgent \cite{chen2025locagent} at 18 times lower API cost, and an MRR of 0.851 with a Top-1 accuracy of 78.6\% using an open-source model.

\noindent\textbf{RQ2. How does such a pipeline influence end-to-end Automated Program Repair (APR)?}

When integrated into \textsc{Agentless} \cite{xia2025agentless}, an established open-source APR framework, \textsc{BLAgent} improves the overall issue resolution rate by over 20\%, demonstrating that more precise file-level localization directly translates into measurable gains in end-to-end patch synthesis.







\section{BLAgent Methodology}
\label{sec:agentic-localization}



\begin{figure}[htbp]
    \centering
    \includegraphics[width=0.7\linewidth]{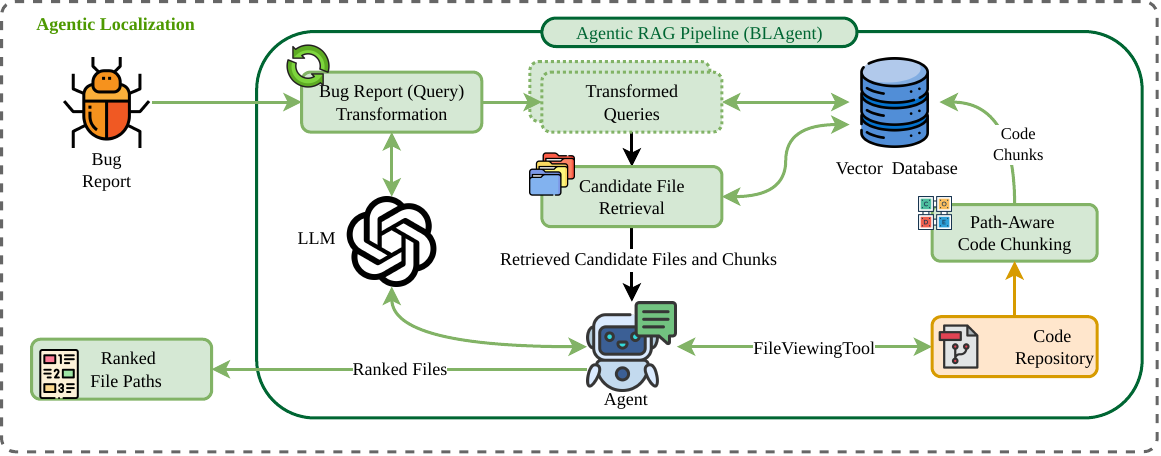}
    \caption{Overall outline of the proposed localization approach.}
    \label{fig:overall-outline}
\end{figure}

We introduce \textsc{BLAgent}, an agentic RAG framework that identifies and ranks source files likely to require modification for a given bug report. As illustrated in Figure~\ref{fig:overall-outline}, \textsc{BLAgent} follows a structured localization pipeline that progressively narrows the repository search space through retrieval and agentic reranking. The framework operates in four major steps:

\begin{enumerate}
    \item \textbf{Repository Encoding (Section \ref{sec:repo-encoding}).} \textsc{BLAgent} first preprocesses the target repository by extracting source files and segmenting them into syntactically meaningful code chunks using AST-aware splitting. Each chunk is augmented with its relative file path before being embedded and stored in a vector database, enabling retrieval that captures both code semantics and repository structure.

    \item \textbf{Query Transformation (Section \ref{sec:query-transformation-method}).} Given a bug report, \textsc{BLAgent} rewrites it into two complementary retrieval-oriented queries. The structural query emphasizes code entities such as modules, classes, functions, and traceback cues, while the behavioral query captures the observed failure and expected behavior.

    \item \textbf{Candidate File Retrieval (Section \ref{sec:candidate-file-retrieval}).} The transformed queries are independently used to retrieve relevant code chunks from the vector database. Chunk-level similarity scores are aggregated into file-level scores, and the top files (ordered by scores) from both query perspectives are merged into a compact candidate set.

    \item \textbf{Agentic Reranking (Section \ref{sec:agentic-reranking-design}).} Finally, \textsc{BLAgent} reranks the retrieved candidate files through a two-phase agentic process. A ReAct-based agent first inspects file skeletons and assigns relevance scores, after which an evidence-anchored reranking step expands only retriever-highlighted code regions to produce the final ranked file list.
\end{enumerate}

\subsection{Repository Encoding}
\label{sec:repo-encoding}
Given a target repository, we begin with repository-level preprocessing and encoding. All source files (e.g., \texttt{.py}) are extracted, while non-executable files (e.g., licenses, markdowns) are filtered out since they rarely contribute to bug fixing. The selected source files are then segmented into smaller, semantically meaningful units (i.e., chunks) for vector encoding and storage. Instead of relying on naive text-based chunking—where code is divided by arbitrary character or line limits—we employ a \emph{code chunking} strategy using \texttt{CodeSplitter}\footnote{\url{https://developers.llamaindex.ai/python/framework-api-reference/node_parsers/code/\#llama_index.core.node_parser.CodeSplitter}}. This approach leverages the abstract syntax tree (AST) of the target language to produce chunks that align with syntactic and semantic boundaries (e.g., functions, classes, or logical code blocks). Such segmentation aims to preserve the structural and semantic integrity of the chunks.

\begin{figure}[h!]
\centering
\begin{minipage}[t]{0.47\textwidth}
\centering
\caption*{\textbf{(a) Naive text-based chunking}}
\begin{lstlisting}[style=code]
def compute_average(nums):
    total = sum(nums)
    count = len(nums)
    return total / count

def normalize(nums):
    mean = compute_average(nums)
    range_val = max(nums) - min(nums)
    return [(n - mean) / range_val

###### SPLIT ######

for n in nums]

def scale(nums, factor):
    return [n * factor for n in nums]
...
\end{lstlisting}
\end{minipage}
\hfill
\begin{minipage}[t]{0.47\textwidth}
\centering
\caption*{\textbf{(b) Code-aware w/ file path chunking (AST-based)}}
\begin{lstlisting}[style=code]
[PATH] src/core/utils/math.py
[CODE]
def compute_average(nums):
    total = sum(nums)
    count = len(nums)
    return total / count

###### SPLIT ######

[PATH] src/core/utils/math.py
[CODE]
def normalize(nums):
    mean = compute_average(nums)
    range_val = max(nums) - min(nums)
    return [(n - mean) / range_val for n in nums]
...
\end{lstlisting}
\end{minipage}
\caption{Comparison of naive text-based versus AST-aware code splitting. 
The naive splitter (a) breaks the \texttt{normalize()} function mid-expression, 
while the code-aware splitter (b) maintains full structural integrity.}
\label{fig:split-example}
\end{figure}

\noindent\textbf{\rev{Semantic Preservation.}} \rev{We define \emph{semantic preservation} as ensuring that no chunk cuts across an AST construct. For example, splitting a function body mid-expression or separating a decorator from its decorated function. Such fragmentation strips away local dependencies (e.g., parameter bindings, loop invariants, return contracts) that the embedding model requires to produce a meaningful vector representation. As shown in Figure~\ref{fig:split-example}, naive text splitting breaks the \texttt{normalize()} function mid-expression, merging incomplete fragments into a single incoherent chunk, whereas AST-aware splitting retains each function as a self-contained logical unit. When a function exceeds the maximum chunk size, the splitter recursively subdivides it at the finest available AST boundaries---such as nested blocks, conditionals, or statement-level nodes---rather than at an arbitrary character offset. Each resulting sub-chunk thus still represents a syntactically complete program fragment, preserving local variable scopes and control-flow context within that segment.}

\noindent\textbf{\rev{File Path Augmentation.}}
\rev{To further enhance contextual precision, each chunk is augmented with its \emph{file 
path} (e.g., \texttt{src/utils/math.py}). This augmentation creates a direct alignment channel between bug reports and code chunks. For example, bug reports  frequently contain implicit or explicit \textit{path-like} signals---module references (e.g., \textit{django.db.models.fields}), traceback entries (e.g., \textit{File "django/db/migrations/writer.py"}), or package-qualified identifiers (e.g., \textit{sklearn.linear\_model.LogisticRegressionCV})---that correspond structurally to file system paths. Without path augmentation, these signals are present in the query embedding but absent from the chunk embedding, resulting in a systematic vocabulary mismatch. Prepending the relative file path to each chunk closes this gap: the embedding model jointly encodes content and location; therefore, path-bearing tokens in the bug report directly reinforce similarity with the chunks they describe. This is also useful for identically named entities across different project hierarchies (e.g., \texttt{module/classA.method\_x} vs. \texttt{module/classB.method\_x}), where content-only embeddings would otherwise assign near-identical similarity scores to structurally unrelated functions. Figure \ref{fig:split-example} illustrates the difference between naive text chunking and our code-aware, path-augmented chunking. The final chunks are embedded using a pre-trained model to obtain dense vector representations, stored in a vector database for efficient similarity-based retrieval. We adopt \emph{dense retrieval}, as prior work demonstrates its superiority over traditional sparse methods~\cite{sawarkar2024blended, parvez2021retrieval}.}

\rev{However, we did not consider LLM-based semantic chunking, where a language 
model identifies topically coherent boundaries and segments code accordingly. Recent work 
suggests that the advantages of semantic chunking are highly task-dependent and frequently fail to justify the added computational cost ~\cite{qu2025semantic}, whereas AST-based chunking yields self-contained, semantically coherent units that improve retrieval on code-related tasks~\cite{zhang2025cast}. Given that our experimental setup spans the full SWE-bench-Lite benchmark---repositories averaging over 11,000 functions each---applying such chunking involving an LLM at indexing time would incur substantial cost that is difficult to justify.}


\subsection{Query Transformation}
\label{sec:query-transformation-method}
\begin{figure}[htbp]
    \centering
    \begin{minipage}{\linewidth}  
        \begin{lstlisting}[style=prompt, caption={Structural query transformation prompt (PT$_0$).}, label={lst:structural_augmentation_prompt}]
You are an AI assistant that rewrites Python bug reports and technical descriptions into focused search queries to retrieve source code most relevant to the root cause of the issue.

Focus on the code structure and error source. Include:
- Relevant functions, methods, or classes (e.g., `__init__`, `from_pretrained`, `__repr__`)
- Module or package names (e.g., `sklearn.linear_model`, `torch.nn.functional`)
- File or traceback context if available (e.g., error in `sklearn/utils/validation.py`)
- Probable reason (e.g., wrong type comparison, improper shape validation, missing null check)
- Avoid restating full tracebacks or irrelevant details
- Do not include user-defined variable/class names, or specific instance details unless they are part of the core issue.

### Example
Bug:
    With `sklearn.set_config(print_changed_only=True)`, printing `LogisticRegressionCV(Cs=np.array([0.1, 1]))` raises:
    ValueError: The truth value of an array with more than one element is ambiguous.

Transformed:
    In `sklearn.linear_model.LogisticRegressionCV`, enabling `print_changed_only=True` causes a ValueError when `Cs` is a numpy array. The issue likely stems from a parameter comparison in `__repr__` or `__init__` using `!=` on arrays without `.all()` or `.any()`. Suspected faulty logic in parameter diffing or config-aware repr code in `sklearn.utils._param_validation` or related helpers.

Output only the rewritten query text. No sections, headers, or markdowns. Be retrieval-friendly and concise, with no extra text. Aim for 2-4 sentences.
\end{lstlisting}
    \end{minipage}
\end{figure}
Directly using the raw text of a bug report for retrieval can lead to suboptimal results as lexical mismatches between natural language descriptions and code identifiers, as well as extraneous narrative or contextual information, can distort query embeddings and hinder retrieval of relevant code fragments \cite{li2024dmqr, shao2024enhancing, xiao2019improving}. In practice, relevant code entities may be described either structurally (e.g., via class or method names) or behaviorally (e.g., through observed runtime behavior), but these aspects are not always explicitly stated in the surface text.

To address this challenge, we employ a query transformation approach inspired by retrieval-augmented generation (RAG) methods (e.g., rewriting, decomposition, etc.) that improve document-query alignment \cite{li2024dmqr, chan2024rq, ma2023query}. The key idea is to reformulate a bug report into retrieval-oriented queries that emphasize complementary perspectives of the fault, bridging the semantic gap between natural language descriptions and the underlying code.

Specifically, we decompose every bug report into two complementary transformations: a structural transformation ($T_0$) and a behavioral transformation ($T_1$). We chose these two because bug reports typically contain both (1) explicit or implicit references to program entities (e.g., function names, APIs, parameters), and (2) descriptions of the erroneous behavior or execution context. The structural transformation ($T_0$) distills the report into its code-related components like identifiers, modules, and relevant structural cues, making it suitable for retrieving code elements that directly match these terms. In contrast, the behavioral transformation ($T_1$) reformulates the report, capturing how the system is expected to behave versus how it actually behaves.

\begin{figure}[htbp]
    \centering
    \includegraphics[width=\linewidth]{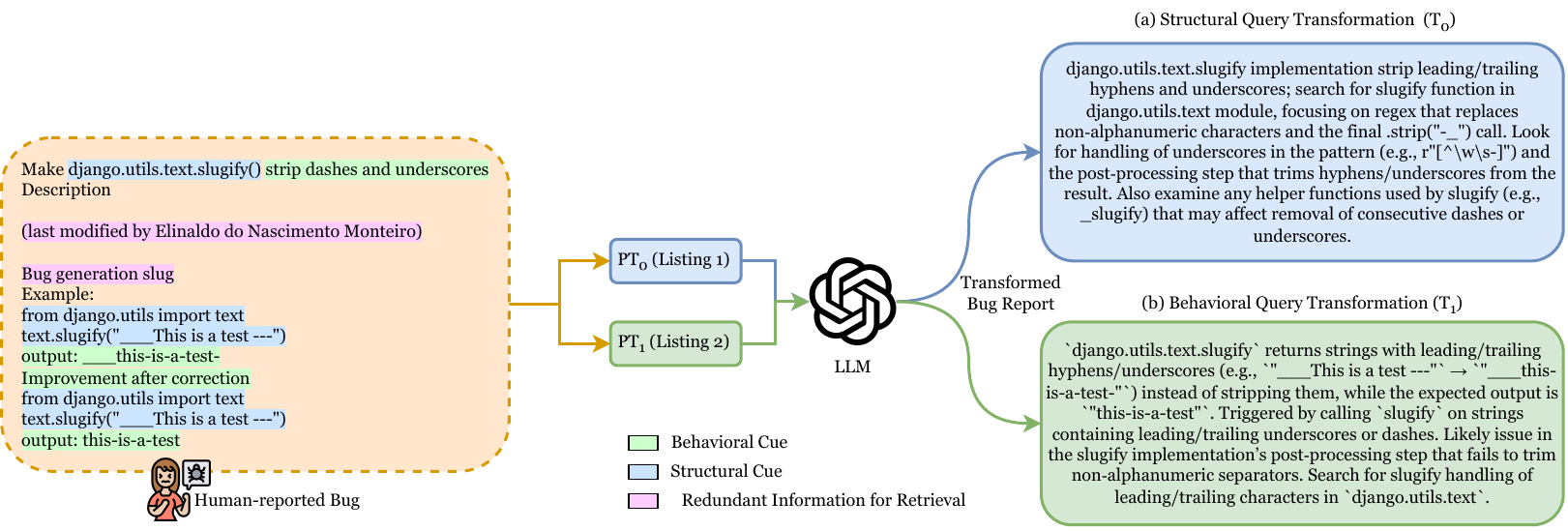}
    \caption{Query transformation of human-reported bug (Example Bug: \texttt{django\_\_django-12983}).}
    \label{fig:query-transformation}
\end{figure}
\noindent\textbf{Structural Query Transformation $(T_0)$.}
This transformation focuses on the static, syntactic aspects of the bug report, extracting signals related to identifiable code entities—such as modules, functions, and helper routines. We use prompt $(PT_0)$ in Listing \ref{lst:structural_augmentation_prompt} to transform queries for structural augmentation. We provide one example to the LLM to show how this can be done. In Listing \ref{lst:structural_augmentation_prompt}, the example shows that when a specific configuration is set, initializing \texttt{LogisticRegressionCV(...)} ends up in an error. The transformed query also adds more structural cues like \texttt{\_\_init\_\_}, \texttt{\_\_repr\_\_} since the issue happens during initialization. This is particularly beneficial for dense retrieval where reasoning is not available and its complete dependency on the embedding similarity of query and the documents.

Similarly, we provide one real example shown in Figure \ref{fig:query-transformation}a, where the transformation focused on lexical and structural components. The transformed query explicitly names the relevant function (\texttt{slugify}), module (\texttt{django.utils.text}), possible helper routine (\texttt{\_slugify}), and even points to specific syntactic constructs such as the regular expression and post-processing call (\texttt{.strip()}). This structural transformation of the original bug guides retrieval toward locations in the repository where the implementation logic or post-processing behavior may reside.


\begin{figure}[t]
    \centering
    \begin{minipage}{\linewidth}  
        \begin{lstlisting}[style=prompt, caption={Behavioral query transformation prompt (PT$_1$).}, label={lst:bhavioral_augmentation_prompt}]
You are an AI assistant that rewrites Python bug reports and behavioral descriptions into focused search queries for retrieving the most relevant source code responsible for the issue.

Focus on the user-observed behavior and triggering conditions. Include:
- The exact observed behavior (e.g., silent failure, incorrect output, crash)
- Expected vs. actual behavior
- Triggers such as input parameters, environment, CLI flags, or API calls
- Probable module or component involved (if clearly inferable)
- Possible reason behind the behavior (e.g., missing fallback, incorrect state check)
- Avoid speculative deep internals; keep it grounded in externally visible symptoms

### Example
Bug:
    When training a model using `transformers.Trainer` with `fp16=True` on a 4GB GPU, training silently hangs. No traceback or error message is shown; the training loop remains stuck after the first step.

Transformed:
    `transformers.Trainer` hangs during training when `fp16=True` and VRAM is low (e.g., 4GB GPU). Likely module involved: `accelerate` or mixed precision handling in `Trainer`. Expected OOM error or graceful fallback, but instead the training loop freezes without exception, possibly due to unhandled CUDA error or silent failure in gradient scaler.
    
Output only the rewritten query text. No sections, headers, or markdowns. Be retrieval-friendly and concise, with no extra text. Aim for 2-4 sentences.
\end{lstlisting}
    \end{minipage}
\end{figure}

\noindent\textbf{Behavioral Query Transformation $(T_1)$.} The behavioral query transformation specifically focuses on the runtime behavior described in the bug report, capturing cues about faulty execution, unexpected outputs, or missing functionality. As shown in Figure~\ref{fig:query-transformation}b, this transformation reframes the bug in terms of observed program behavior and expected outcomes, allowing the retriever to better connect textual failure descriptions to the underlying runtime logic.

For example, in the \texttt{slugify} case, the raw bug report describes an undesired output string produced by the function (i.e., it fails to strip leading and trailing hyphens or underscores). The behavioral transformation reformulates this report to explicitly contrast the observed behavior (“returns strings with trailing underscores or dashes”) with the expected outcome (“should strip non-alphanumeric separators”). This reframing highlights the symptom in a way that points the retriever toward code regions responsible for normalization or post-processing. We use prompt $(PT_1)$ (Listing~\ref{lst:bhavioral_augmentation_prompt}) to elicit such transformations, guiding the LLM to emphasize runtime mismatches, incorrect outputs, and failed handling of edge cases.


\subsection{Candidate File Retrieval}
\label{sec:candidate-file-retrieval}
Let $T_0$ and $T_1$ denote the two query transformations (structural and behavioral, respectively). For a transformed query $q^{(k)}$ with $k\in{0,1}$, we embed the query using the same encoder applied to repository chunks. Each repository file $f$ is partitioned into chunks, and each chunk $c_{i,j}$ (the $j$-th chunk of file $f_i$) has an embedding $\mathbf{c}_{i,j}$ produced by the same encoder. Similar code chunks are then retrieved from the database using nearest neighbour search \cite{malkov2018efficient}.
As a code file typically contains multiple functions or logical units while a bug is most often localized to a single unit, we aggregate chunk-level similarity scores for each file by taking the maximum over its chunks. The file-level similarity under query transformation $T_k$ is therefore
\[
\mathrm{sim}\big(q^{(k)}, f_i\big) = S^{(k)}*i
= \max*{j \in C(f_i)} s^{(k)}_{i,j},
\]
where $C(f_i)$ is the index set of chunks in file $f_i$. Taking the maximum highlights the single region of a file that best matches the query and avoids diluting strong local matches by averaging with unrelated code.

For each transformation $(T_k)$ we retrieve a ranked list of files ordered by decreasing file-level similarity:
\[
{FT_k} = \big[ f^{(k)}*1, f^{(k)}*2, \dots, f^{(k)}*{n_k} \big],
\qquad
\mathrm{sim}\big(q^{(k)}, f^{(k)}*i\big) \ge \mathrm{sim}\big(q^{(k)}, f^{(k)}*{i+1}\big).
\]
To form the agent’s candidate input we take the top-(m) files from each list, concatenate these two prefixes, and remove duplicates while preserving first-occurrence order:
\[
{F_\text{candidate}} = \mathrm{unique}\Big( {FT_0}^{[:m]} \oplus {FT_1}^{[:m]} \Big),
\]
where $(\oplus)$ denotes concatenation, ${FT_k}^{[:m]}$ represents the first $m$ files from the ranked list $FT_k$, and $\mathrm{unique}(\cdot)$ eliminates repeated file entries while keeping the ordering induced by the concatenation. By construction each file appears at most once in ${F}_{\text{candidate}}$. We limit the number of candidates to 15. \rev{Although this concatenation order places $T_0$ candidates first by default, Section~\ref{sec:rq1-4} shows that reversing the order produces negligible difference in localization accuracy, confirming that the ordering is an implementation convenience rather than a structural dependency.}

Finally, the agent receives the candidate file set ${F}_{\text{candidate}}$ including the chunks in each candidate together with the original problem statement for downstream reasoning and reranking.

\subsection{Agentic Reranking} 
\label{sec:agentic-reranking-design}

\begin{figure}[htbp]
    \centering
    \includegraphics[width=0.95\linewidth]{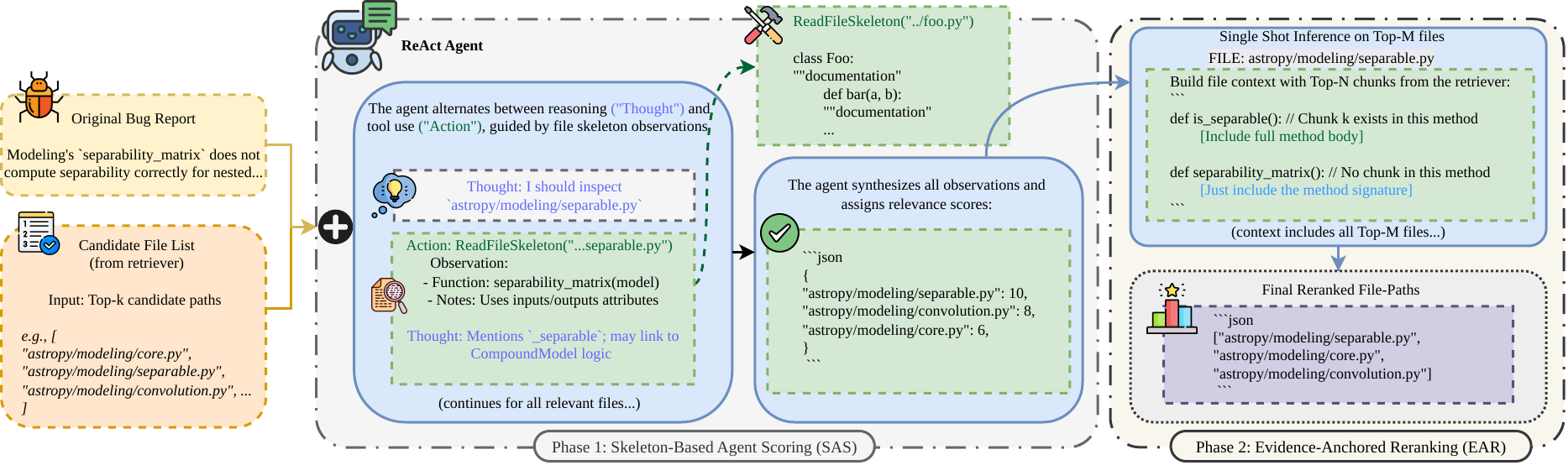}
    \caption{Reranking of the candidate files with ReAct agent.}
    \label{fig:agentic-rag}
\end{figure}

Figure~\ref{fig:agentic-rag} illustrates how \textsc{BLAgent} reranks the candidate file set $F_{\text{candidate}}$ obtained from the retrieval stage. The reranking proceeds in two phases. In Phase 1, a reasoning agent inspects structural skeletons of candidate files and assigns relevance scores. In Phase 2, a one-shot LLM inference reranks the top-scored files using pruned, chunk-grounded file contexts to resolve near-ties and verify implementation-level relevance.

\begin{figure}[htbp]
    \centering
    \begin{minipage}{\linewidth}  
        \begin{lstlisting}[style=prompt, caption={Skeleton-Based Agent Scoring (SAS) Prompt.}, label={lst:reranking_prompt}]
You are a powerful AI code assistant.

Bug Report:
{problem_statement}

You are provided with a list of candidate code files that may be relevant to this bug.

Your task:
1. Use the `ReadFileSkeleton` tool to inspect the structure of each file.
2. Assign a relevance score from 0 (not relevant) to 10 (definitely needs to be modified) to each file.
   The score should reflect how likely the file is to require changes to fix the bug described.
3. You MUST view the file skeletons before assigning scores.
4. Construct a dictionary mapping file paths to their relevance scores.
5. You MUST RETURN scores for at least {num_files} files.

MUST NOTE:
- Multiple files can have the same score
- Should provide the complete file paths as they appear for the `ReadFileSkeleton` tool.
- The provided file paths are sorted by the chunk similarity scores, but AVOID making assumptions based on order.

STOP CONDITION:
- After you come to conclusion, you must immediately output the final answer in the required JSON format and STOP.
- Do NOT write another Thought: after receiving the final Observation and the last thing in your output must be:
Final Answer
{{...}}

FORMAT INSTRUCTIONS:
You must respond using this ReAct format exactly:

Thought: I need to inspect the structure of fileA.py
Action: ReadFileSkeleton
Action Input: "fileA.py"
Observation: [skeleton output here]
Final Answer:
```json
{{
    "fileB.py": 8,
    "fileA.py": 7,
    "fileC.py": 7
}}
Your candidate files:
{retrieved_file_paths}
```
\end{lstlisting}
    \end{minipage}
\end{figure}

\noindent\textbf{Phase 1: Skeleton-Based Agent Scoring (SAS).} We instantiate a ReAct-based \cite{yao2022react} reasoning agent that iteratively assesses the retrieved candidates. The agent operates through an explicit reasoning loop (\textit{Thought $\rightarrow$ Action $\rightarrow$ Observation $\rightarrow$ Answer}), where each iteration involves (i) forming a hypothesis about which files are likely relevant, (ii) selectively inspecting file representations through the equipped tool \texttt{ReadFileSkeleton}, and (iii) updating its internal assessment before producing a final scored output. This reasoning pattern allows the agent to dynamically decide which files to 
examine more closely based on observed evidence, rather than relying solely on static similarity scores.

To facilitate symbolic inspection, each code file is represented as a structural \emph{skeleton} capturing its classes, function signatures, and docstrings. Rather than processing full source code, the agent operates on these skeletal representations to identify files most relevant to the reported bug. For file-level localization, the most informative cues are typically structural and declarative---function names, signatures, and docstrings---rather than implementation details. This abstraction enables the agent to reason efficiently about code organization while staying within a manageable context budget.

The agent is prompted to assign a relevance score in the range of 0--10 (not relevant to highly relevant) to each candidate file (Listing~\ref{lst:reranking_prompt}), reflecting how likely it is to require modification to resolve the reported bug. By iteratively inspecting file skeletons, it refines its confidence estimates and produces a final mapping from file paths to scores.

\begin{figure}[htbp]
    \centering
    \begin{minipage}{\linewidth}  
        \begin{lstlisting}[style=prompt, caption={Evidence-Anchored Reranking (EAR) Prompt.}, label={lst:reranking_prompt_phase2}]
The following are the top {len(file_paths)} ranked files retrieved for a given bug report. Analyze and rerank them based on their relevance to the problem statement. The idea is to find the actual file where a patch needs to be applied to fix the bug.

Return a ranked list of file paths, ordered from most relevant to the least, based on their content and relevance to the problem statement.
Example Output:
```json
{{
    "ranked_files": [
        "path/to/most_relevant_file.py",
        "path/to/second_most_relevant_file.py",
        ...
        "path/to/least_relevant_file.py"
    ]
}}
```
Do not include any explanations or additional text outside the JSON structure.

Problem Statement:
{problem_statement}
Possibly Relevant Files:
{aggregated_code_text}
\end{lstlisting}
    \end{minipage}
\end{figure}

\noindent\textbf{\rev{Phase 2: Evidence-Anchored Reranking (EAR).}}
\rev{Skeleton-based scoring provides an efficient, organization-level view of each candidate file. To further refine this ordering with implementation-level evidence---particularly when multiple candidate files implement similar interface or inherit from a common base class, making their skeletons structurally indistinguishable at the signature level---we introduce a one-shot evidence-anchored reranking step over the top-$M$ Phase~1-scored files. We set $M=5$ to restrict EAR to a small set of highly plausible candidates, enabling effective joint comparison while keeping the reasoning context bounded. This design choice is further supported by our ablation study (Section \ref{sec:context-size-effect}), which shows diminishing returns when increasing the number of candidates. Similarly, for each of the $M$ files, we construct a \emph{pruned file context} using the retriever's top 5 chunks as evidence anchors: we preserve global file structure (imports, class definitions, and method/function signatures) while expanding only the methods that contain retrieved chunks to their full implementations; all other methods are kept as signatures. This exposes implementation-level detail precisely where the retriever considers it most relevant, without loading entire files. The LLM jointly compares these $M$ pruned contexts in a single inference pass using the prompt defined in Listing \ref{lst:reranking_prompt_phase2} and outputs the final ranked list of file paths, improving ranking precision while controlling context length and inference cost.}

\rev{This two-phase design is deliberate. Pruned-context reranking 
over all 15 candidates would significantly inflate context size, increasing cost and risking mid-context attention degradation~\cite{liu2023lost}; moreover, the ReAct agent contributes an independent localization signal through hypothesis-driven structural inspection that no single static inference pass can replicate. Phase~1 thus both reduces the candidate set to a tractable size and enriches the ranking signal that Phase~2 refines.}

\section{Accuracy Assessment of Bug Localization in BLAgent (RQ1)}

To evaluate how effectively \textsc{BLAgent} localizes bugs at the file level, we decompose RQ1 into the following sub-questions:
\begin{itemize}
    \item \textbf{RQ1.1: Can \textsc{BLAgent} outperform state-of-the-art baselines?}  
    
    We investigate whether agentic reasoning in a RAG pipeline (i.e., \textsc{BLAgent}) can improve file-level localization beyond current state-of-the-art methods.     
    
    \item \textbf{RQ1.2: How do different retrieval and RAG configurations influence file-level localization accuracy?}

    We evaluate different chunking and retrieval configurations in a traditional RAG pipeline to identify the most effective setup for file-level localization.
    
    \item \textbf{RQ1.3: Can agentic reranking improve over traditional RAG?}
    
    We examine if agentic reranking can improve over traditional RAG pipelines in localization by using a ReAct-based agent to rerank dense retrieval candidates.
    
    \item \textbf{RQ1.4: Can query transformation further improve localization?}
    
    We investigate whether reformulating bug reports into structured, retrieval-oriented queries further improves localization.

    \item \textbf{RQ1.5: Does the choice of LLM variant affect the performance of the agent?}
    
    We analyze whether the underlying LLM influences the agent’s ability to accurately localize faults. Specifically, we examine how different LLM variants—varying in size and type (e.g., closed or open-source) impact the overall effectiveness.

    \item \textbf{\rev{RQ1.6: Can \textsc{BLAgent} be extended to function-level localization?}}

    \rev{We assess whether the proposed architecture can be generalized to function-level localization.}
\end{itemize}

\subsection{Experimental Setup}
\subsubsection{Localization Baselines.} 
\label{sec:paper-baselines}
We compare \textsc{BLAgent} against recent dedicated bug localization frameworks and localization components within APR systems. Among dedicated localization frameworks, \rev{LocAgent \cite{chen2025locagent} adopts a graph-guided LLM agent framework that represents code repositories as directed heterogeneous graphs encoding structural elements and dependencies, enabling multi-hop reasoning to navigate and localize relevant code entities efficiently.} CoSIL \cite{jiang2025cosil} introduces an LLM-driven, function-level localization framework that performs a two-phase code graph search — first exploring broadly at the file level through module call graphs, then refining at the function-level with pruning and reflection mechanisms to control search direction and context quality. BugCerberus \cite{chang2025bridging} adopts a hierarchical localization design powered by specialized LLMs operating at the file, function, and statement levels, which progressively narrow down bug locations by leveraging intermediate program representations and contextual cues.

\rev{Among end-to-end APR frameworks, Agentless \cite{xia2025agentless} combines both an embedding-based retrieval method and an LLM-based approach that reasons over the project's file tree to rank relevant files. AutoCodeRover \cite{zhang2024autocoderover} is an agentic framework that utilizes multiple specialized agents alongside AST-based code search APIs and spectrum-based fault localization to iteratively localize buggy code and propose fixes. We also compare the localization accuracy with OpenHands \cite{wang2024openhands}, which uses its default CodeAct agent that localizes bugs by iteratively issuing shell and code execution actions to explore the repository.}

\subsubsection{Evaluation Metrics}
Following prior work~\cite{chang2025bridging, jiang2025cosil, xia2025agentless}, we employ standard information retrieval metrics to assess localization performance.

\noindent\textbf{Mean Reciprocal Rank (MRR).}  
MRR measures the average of the reciprocal ranks of the first correctly localized file for each bug report. Formally, let $Q$ denote the set of bug reports, and $r_i$ the rank position of the first correctly localized file for query $i$. Then,
\begin{equation}
\mathrm{MRR} = \frac{1}{|Q|} \sum_{i=1}^{|Q|} \frac{1}{r_i}.
\end{equation}
A higher MRR indicates that the model ranks the correct file closer to the top.

\noindent\textbf{Top-$k$ Accuracy.}  
Top-$k$ Accuracy evaluates the proportion of bug reports for which at least one correct bug location appears within the top-$k$ ranked predictions. Let $\mathbb{1}[\cdot]$ denote the indicator function, which equals 1 if the condition is true and 0 otherwise. Then,
\begin{equation}
\mathrm{Top\text{-}k} = \frac{1}{|Q|} \sum_{i=1}^{|Q|} \mathbb{1}[\text{correct file} \in \text{Top-}k_i].
\end{equation}
We report Top-$\{1,3,5,10\}$ accuracy to evaluate the competitive methods. 

\subsubsection{Dataset}
For our experiments, we use the \textit{SWE-bench-Lite} dataset \cite{jimenez2024swebench}, which comprises 300 real-world bug instances collected from open-source Python projects. Each entry in the dataset contains a detailed bug report, the corresponding commit hash of the repository representing the HEAD of the repository before the issue was fixed, ground truth patch, etc. SWE-bench-Lite is specifically designed for evaluating bug localization and program repair techniques, providing a diverse set of bugs that vary in complexity, type, and affected code components.

\subsubsection{Large Language Models.}
We use the \texttt{GPT-OSS-120B} \footnote{\url{https://openai.com/index/introducing-gpt-oss/}} model from Ollama as the primary LLM for all agentic reasoning and text-generating components in our framework and the downstream APR framework. The model uses 128 mixture-of-experts (MoE) \cite{shazeer2017outrageously} and has a context length of 128,000. We particularly adopt this model due to its openness and proven reasoning capability compared to other proprietary models like GPT-o4-mini, or o3-mini on different benchmarks. \rev{Additionally, we also validate our localization results with SOTA proprietary LLM---\texttt{Claude-4.6-Sonnet (Claude-4.6)}.}

\subsubsection{Embedding Model.}
For text and code representation, we employ the \texttt{nomic-ai/nomic-embed-text-v1} model\footnote{\url{https://huggingface.co/nomic-ai/nomic-embed-text-v1}}
 to generate high-quality embeddings used in retrieval and similarity-based ranking. This model was selected for its strong empirical performance—comparable to proprietary alternatives such as OpenAI’s text-embedding-3-small—and its large context window of 8192 tokens, enabling effective encoding of longer code segments and detailed bug reports.


\subsubsection{System Configuration} All experiments were conducted on a system equipped with 2$\times$NVIDIA L40S 48GB GPUs, an Intel Xeon Gold 6442Y$\times$96 processor, and 250 GB RAM.

\subsection{RQ1.1 Can \textsc{BLAgent} outperform state-of-the-art baselines?}
\label{sec:agentic-vs-existing}



\subsubsection{Approach.}
To assess the effectiveness of the proposed agentic RAG framework, we conduct a comparative evaluation against several state-of-the-art bug localization approaches discussed in Section \ref{sec:paper-baselines}. We evaluate \textsc{BLAgent} using GPT-OSS-120B as the base LLM and \texttt{nomic-embed-text-v1} as the embedding model. \rev{Rather than reproducing all baseline results---which would incur prohibitive API costs---we report each baseline at its best published configuration on SWE-bench-Lite. For systems primarily designed as end-to-end APR pipelines that do not explicitly report file-level localization accuracy, we use results from published papers on the same benchmark. AutoCodeRover results are taken from Chang et al.~\cite{chang2025bridging}, and OpenHands results are taken from Chen et al.~\cite{chen2025locagent}.}

\rev{To further disentangle the contribution of the \textsc{BLAgent} architecture from the choice of LLM, we conduct a controlled comparison where both \textsc{BLAgent} and LocAgent (the strongest Top-$k$ baseline) are run under the same model, Claude-4.6-Sonnet. Due to the substantial API cost of running full benchmark evaluations with proprietary models, we restrict this controlled comparison only to these two systems. We allocate a fixed evaluation budget of \$300 for LocAgent using Claude-4.6-Sonnet to control evaluation cost. Under this constraint, LocAgent processes 182 of the 300 instances before exhausting the budget, and results are reported over this subset (see Section~\ref{sec:cost-analysis} for detailed cost analysis).}

\begin{table}[htbp]
\centering
\caption{File-level localization accuracy of different methods. $^\dagger$ indicates the approaches we reproduced. The baselines include both dedicated localization methods (BugCerberus, CoSIL, LocAgent) and localization methods within end-to-end APR frameworks (Agentless, AutoCodeRover, OpenHands).}
\label{tab:file_level_accuracy}
\small
\begin{tabular}{llccccc}
\toprule
\textbf{Dataset} & \textbf{Approach} & \textbf{MRR} & \textbf{Top-1} & \textbf{Top-3} & \textbf{Top-5} & \textbf{Top-10} \\
\midrule
\multirow{14}{*}{\textbf{SWE-bench-Lite}}
& \multicolumn{6}{l}{\textit{(a) Comparison against published baselines}} \\
\cmidrule(lr){2-7}
 & BugCerberus & 0.733 & 0.651 & 0.745 & 0.754 & 0.791 \\
 & CoSIL (Qwen-2.5-32B) & 0.701 & 0.613 & 0.780 & 0.837 & -- \\
 & LocAgent (Qwen-2.5-32B FT)  & -- & 0.759 & 0.905 & 0.927 & --  \\
 & LocAgent (Claude-3.5) & -- & 0.777 & 0.919 & 0.941 & -- \\
 & Agentless (GPT-OSS)$^\dagger$ & 0.719 & 0.623	& 0.817	& 0.850 & 0.857 \\
 & Agentless (GPT-4o) &0.715 & 0.630 & 0.817 & 0.850 & 0.883 \\
 & AutoCodeRover (GPT-4o) & 0.650 & 0.557 & 0.698 & 0.741 & 0.754 \\
 & OpenHands (GPT-4o) & -- & 0.609 & 0.719 & 0.737 & -- \\
 & OpenHands (Claude-3.5) & -- & 0.762 & 0.897 & 0.901 & -- \\
 & \textbf{\textsc{BLAgent} (GPT-OSS)} & 0.851 & 0.786 & 0.923 & 0.933 & 0.943 \\
 \cmidrule(lr){2-7}
 & \multicolumn{6}{l}{\textit{(b) Controlled comparison under the same LLM}} \\
 \cmidrule(lr){2-7}
 & LocAgent (Claude-4.6)$^\dagger$ & -- & 0.824 & 0.863 & 0.863 & -- \\
& \textbf{\textsc{BLAgent} (Claude-4.6)} & \textbf{0.900} & \textbf{0.867} & \textbf{0.930} & \textbf{0.947} & \textbf{0.953} \\
\bottomrule
\end{tabular}
\end{table}
\subsubsection{Results} 
Table~\ref{tab:file_level_accuracy} reports the overall file-level localization accuracy for \textsc{BLAgent} (Phase 1 + Phase 2) compared to leading baselines on the SWE-bench-Lite dataset. The results indicate a consistent advantage for agentic RAG across all metrics.

\noindent\textbf{Comparison against dedicated localization methods.} \textsc{BLAgent} (GPT-OSS-120B) achieves MRR 0.851 with a Top-1 accuracy of 78.6\%, outperforming all baselines at their best published configurations. LocAgent (Claude-3.5) is the closest competitor, achieving Top-1 77.7\% and Top-3 91.9\%, yet \textsc{BLAgent} surpasses it on Top-1 and Top-3 while relying on an open-source model and without requiring any specialized fine-tuning, pre-built repository dependency graphs, or static analysis infrastructure. These gains are particularly meaningful at lower $k$ values, as downstream repair systems often restrict consideration to only the top-1 or top-3 ranked files due to cost and context window constraints \cite{xia2025agentless}.

\noindent\textbf{\rev{Comparison against APR-based localization strategies.}} \rev{\textsc{BLAgent} consistently outperforms the localization components used within existing APR systems under standard ranked retrieval metrics. Agentless \cite{xia2025agentless}, combining embedding-based retrieval and LLM reasoning, achieves Top-1 63.0\% with GPT-4o. AutoCodeRover \cite{zhang2024autocoderover}, which performs localization through agentic code search using AST-based APIs and spectrum-based signals, achieves 55.7\% Top-1 accuracy. Among APR systems, OpenHands \cite{wang2024openhands} achieves the strongest file-level localization (76.2\% Top-1 accuracy with Claude-3.5), yet remains below \textsc{BLAgent} (78.6\%) using an open-source model. However, we note that not all APR systems report localization under comparable evaluation protocols. In particular, LingmaAgent \cite{ma2025alibaba} reports a file-localization accuracy of 67.7\% based on a patch-level criterion—checking whether the generated patch targets the correct file—rather than ranked retrieval metrics (Top-$k$, MRR). As such, this result is not directly comparable. Nevertheless, even under a conservative interpretation treating this figure as Top-1 accuracy, it remains below multiple baselines reported in Table~\ref{tab:file_level_accuracy}, including \textsc{BLAgent}.}

\noindent\textbf{\rev{Controlled comparison under the same LLM.}} \rev{When both systems are run under Claude-4.6-Sonnet (Claude-4.6), \textsc{BLAgent} achieves an MRR of 0.900 and a Top-1 of 86.7\%, Top-3 of 93.0\%, and Top-5 of 94.7\% across all 300 instances. The closest competitor, LocAgent (Claude-4.6), evaluated on 182 completed instances, achieves a Top-1 of 82.4\% --- higher than its Claude-3.5 result (77.7\%), suggesting that the stronger model improves Top-1 predictions. However, its Top-3 and Top-5 accuracy both drop to 86.3\%, falling well below its own Claude-3.5 results (91.9\% and 94.1\%, respectively). For a fair comparison, we evaluated the localization results of \textsc{BLAgent} on the same 182 instances, and it demonstrated a 90.6\% Top-1 accuracy. Furthermore, we inspected the localization results of LocAgent and found that Claude-4.6 causes LocAgent to return fewer candidate files in most instances, often only one or two, reflecting overconfidence in its top prediction at the expense of recall. In contrast, \textsc{BLAgent} with Claude-4.6-Sonnet improves consistently across all thresholds, confirming that the architectural design --- bounded agentic reasoning over a structured candidate set --- is the primary driver of its gains rather than the underlying LLM.}

\begin{observation}{obs:localization-performance}
\textsc{BLAgent} outperforms most baseline localization methods. Iterative reasoning combined with contextual generation enables a deeper contextual understanding of bug reports. 
\end{observation}


\subsubsection{Discussion} While \textsc{BLAgent} demonstrates superior performance compared to state-of-the-art methods across all Top-k, it is important to examine the cases where it fails to correctly localize faulty files. While using Claude-4.6, we find that in all 14 failed cases, the correct file was not retrieved in the top 15 candidates by any of the query transformation techniques.

\noindent\textbf{\rev{Failure case analysis.}} \rev{A systematic analysis 
of the 14 retrieval failures reveals three distinct patterns, summarized 
in Table~\ref{tab:retrieval_failure_analysis}. Two cases involve \textit{feature requests}, where no existing code snippet (e.g., function) was modified by the patch, and the fix resides in configuration or infrastructure code with no strong retrieval signal. Notably, our baseline LocAgent explicitly excludes these same instances from its own evaluation, stating they do not modify any existing function \cite{chen2025locagent}. Five cases involve \textit{hidden dependencies}, where retrieval correctly identifies the primary symptom site but the fix resides in an imported utility or base class that the retrieved file depends on. The remaining seven are \textit{vocabulary mismatch} cases, where the bug report's surface text strongly suggests one module while the fix resides in a behaviorally connected but lexically distant one. To investigate the recall ceiling, we extended retrieval up to Top-50 per query transformation for the 12 non-feature-request failures. The correct file appears at ranks 18--50 in 8 instances, while the remaining 4 are absent from the top-50 of both transformations, asserting that an embedding-level semantic gap that merely increasing $k$ may not resolve. However, extending $k$ to 50 inflates the candidate pool ($T_0 + T_1$) from 15 to $\sim$63 unique files on average, incurring substantial cost and context overflow in the agentic reranking pipeline.}

\begin{table}[htbp]
\centering
\caption{Taxonomy of the 14 retrieval failures where the correct file was absent from the Top-15 candidate pool.}
\label{tab:retrieval_failure_analysis}
\small
\begin{tabular}{lp{4cm}p{7cm}}
\toprule
\textbf{Category} & \textbf{Bug ID} & \textbf{Description} \\
\midrule

Feature Request
    & \texttt{django-11564},
      \texttt{sympy-20590}
    & No faulty file exists to localize --- the patch adds new logic 
      to infrastructure or configuration code with no semantic signal 
      in the bug report (e.g., \texttt{django-11564} adds 
      \texttt{SCRIPT\_NAME} support via \texttt{conf/\_\_init\_\_.py} 
      while the report describes template tag behavior). \\
\midrule

Hidden Dependency
    & \texttt{django-14997},
      \texttt{django-16400},
      \texttt{sympy-13146},
      \texttt{sympy-18087},
      \texttt{sympy-21612}
    & Retrieval correctly identifies the symptom site but the fix 
      resides in an imported utility or base class. For example, 
      \texttt{django-14997} retrieves \texttt{sqlite3/schema.py} 
      (the crash site) while the fix is in \texttt{ddl\_references.py} 
      which it imports. \\
\midrule

Vocabulary Mismatch
    & \texttt{django-11797},
      \texttt{sympy-12236},
      \texttt{sympy-13031},
      \texttt{sympy-13915},
      \texttt{sympy-20322},
      \texttt{sympy-21379},
      \texttt{sympy-21627}
    & The bug reports strongly suggest one module while the fix resides in a behaviorally connected but lexically distant one. For example, \texttt{sympy-20322} describes a \texttt{ceiling} inconsistency, directing retrieval to \texttt{integers.py} as it has multiple ceiling and rounding methods, while the fix is in \texttt{core/mul.py}. \\

\bottomrule
\end{tabular}
\end{table}

\noindent\textbf{\rev{Repository-level concentration.}} \rev{Ten of the 14 failures originate from the \texttt{sympy} repository, representing a 13.0\% repository failure rate (10 out of 77 instances), compared to only 3.5\% for \texttt{django} (4 out of 114 instances). This concentration reflects a systematic mismatch between how \texttt{sympy} bug reports are written and how its implementation is organized. In 9 of 10 failed \texttt{sympy} cases, \textsc{BLAgent} retrieves at least one file from the same directory as the true patch, indicating that the relevant neighborhood is usually identified correctly. Bug reports describe the user-visible mathematical symptom, which consistently points retrieval toward the observable behavior rather than the underlying implementation site. For example, in \texttt{sympy-21612} the report describes a ``LaTeX parsing'' error, so retrieval surfaces \texttt{parsing/latex/} and \texttt{printing/latex.py}; yet none of the dominant report terms appear in the actual patch file \texttt{printing/str.py}. In \texttt{sympy-18087}, the terms \textit{trigsimp}, \textit{cos}, and \textit{sqrt} are absent from the patch file \texttt{core/exprtools.py}, making it invisible to embedding-based retrieval. This is compounded by \texttt{sympy}'s mathematical naming conventions: files such as \texttt{exprtools.py} and \texttt{operations.py} implement low-level algebraic primitives that are silently invoked by the higher-level operations users interact with and are described mostly in abstract algebraic terms that bear no lexical resemblance to user-reported symptoms. By contrast, \texttt{django}'s domain-descriptive identifiers align naturally with the terms users write in bug reports, giving the retriever a strong discriminative signal.}



These observations highlight two critical insights: first, dense retrieval remains a crucial component of the pipeline, as it determines the initial pool of files available for reranking; second, improving recall in the retrieval stage may directly translate into improved localization accuracy.

\begin{observation}{obs:localizeation-agentic-performance}
The overall effectiveness of the agentic RAG pipeline is bounded by the recall of its dense retriever. An agent’s reasoning and reranking are only as effective as the candidate pool it receives—if the correct file is missing (e.g., beyond Top-15), subsequent localization becomes unattainable regardless of reasoning capability.
\end{observation}

\subsection{RQ1.2 How do different retrieval and RAG configurations influence file-level localization accuracy?}
\label{sec:effect-of-retrieval}
\begin{figure}[htbp]
    \centering
    \includegraphics[width=0.7\linewidth]{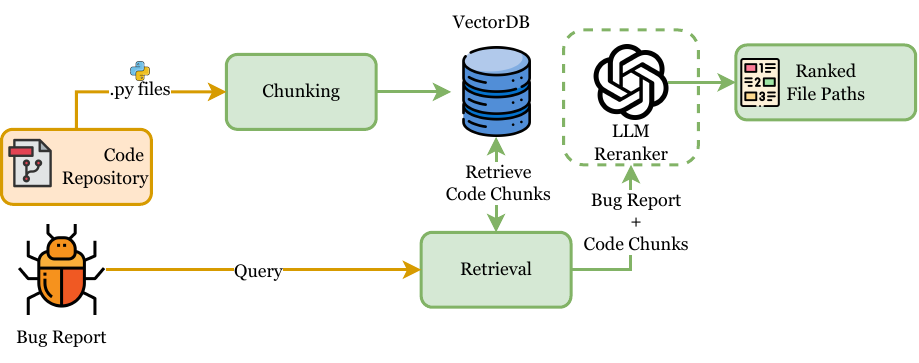}
    \caption{Basic RAG pipeline for file-level localization.}
    \label{fig:basic-rag}
\end{figure}

\subsubsection{Approach.} \rev{We design and evaluate RAG pipelines using different chunking strategies under two ranking settings: (1) Direct Dense Retrieval (Query $\rightarrow$ Retrieval), where candidate files are ranked purely by embedding similarity with no LLM involvement, and (2) LLM Reranking (Basic RAG: Query $\rightarrow$ Retrieval $\rightarrow$ Generation), where the top retrieved chunks are passed as context to an LLM that reranks the candidate files based on their relevance to the bug report.} We extract all source code files (e.g., \texttt{.py} files) from each repository and apply three distinct chunking strategies:
\begin{itemize}
    \item \textit{Text-based chunking}, which splits files into fixed-size textual segments without code structure awareness;
    \item \textit{Code-aware chunking}, which preserves syntactic boundaries (e.g., functions and classes) to retain structural coherence; and
    \item \textit{Code-aware with file path context}, which further embeds each chunk with its file path to provide hierarchical grounding during retrieval (discussed in Section \ref{sec:agentic-localization}).
\end{itemize}
These chunks are then encoded into dense vector representations and stored in a vector database (chromadb\footnote{\url{https://www.trychroma.com/}}) for efficient retrieval. Next, we assess the retrieval accuracy of each chunking strategy independently to measure how effectively relevant code files can be identified given a bug report. Finally, we integrate the retrieval component into a full RAG pipeline (Figure \ref{fig:basic-rag}), where the retrieved code chunks serve as contextual input to an LLM. We provide Top-15 files to the LLM for ranking to keep the context length reasonable. The prompt used for the RAG pipeline is provided in Listing \ref{lst:rag_prompt}. This allows us to find the suitable configuration (e.g., chunking strategy, whether LLM integration is useful or not) that provides the best result.
\begin{figure}[htbp]
    \centering
    \begin{minipage}{0.95\linewidth}  
        \begin{lstlisting}[style=prompt, caption={LLM Prompt for the RAG pipeline.}, label={lst:rag_prompt}]
You are an expert software debugging assistant.

Identify which retrieved files are most likely to contain the bug described below.

Bug Report:
{problem_statement}

Code Context:
{Code Path 1}
{Code Chunk 1}
--
{Code Path 2}
{Code Chunk 2}
...
Instructions:
1. Carefully analyze the bug report and snippets.
2. Respond ONLY with a JSON array of file paths, ordered from most to least relevant.
3. If there is no path provided, return an empty JSON array.
4. DO NOT include any explanations or additional text.
5. Return 10 ranked file paths from the context.
6. DO NOT include paths that were not provided.

Example:
["src/core/utils.py", "src/main/model.py", "src/data/loader.py"]
\end{lstlisting}
    \end{minipage}
\end{figure}

\begin{table}[htbp]
\centering
\caption{Localization accuracy of different retrieval configurations on SWE-bench-Lite. The best setup is marked with $\star$.}
\label{tab:retrieval_accuracy}
\small
\begin{tabular}{l|c|ccccc}
\toprule
\textbf{Chunking} & \textbf{File Ranking} & \textbf{MRR} & \textbf{Top-1} & \textbf{Top-3} & \textbf{Top-5} & \textbf{Top-10} \\
\midrule
\multirow{2}{*}{Text} 
  & Dense Retrieval & 0.459 & 0.350 & 0.526 & 0.600 & 0.686 \\
  & +LLM Reranking   & 0.690 & 0.643 & 0.740 & 0.746 & 0.750 \\
\midrule
\multirow{2}{*}{Code-Aware} 
 & Dense Retrieval & 0.473 & 0.363 & 0.530 & 0.593 & 0.703 \\
 & +LLM Reranking   & 0.722 & 0.667 & 0.783 & 0.793 & 0.800 \\
\midrule
\multirow{2}{*}{Code-Aware w/ File Path $\star$} 
 & Dense Retrieval & 0.553 & 0.417 & 0.627 & 0.713 & \textbf{0.873} \\
 & +LLM Reranking  & \textbf{0.734} & \textbf{0.673} &\textbf{0.796} & \textbf{0.820} & 0.823 \\
\bottomrule
\end{tabular}
\end{table}

\subsubsection{Results.} Table~\ref{tab:retrieval_accuracy} presents the localization performance across different dense retrieval and RAG configurations on the SWE-bench-Lite dataset. The results reveal several important observations.

First, \textit{direct dense retrieval} methods show limited capability in accurately ranking faulty files in top positions. When standard text-based chunking is applied using conventional text-splitters designed for natural language, the performance is notably low. This confirms that code structure and semantics are poorly preserved under text-oriented segmentation. In contrast, \textit{code-aware chunking} slightly improves the MRR and Top-1 accuracy, demonstrating that structural cues such as function boundaries and class definitions contribute to more meaningful embeddings and, consequently, more accurate retrieval. However, we observed that when we embed file paths (e.g., module/src/file.py) along with each chunk, the retrieval performance increases significantly (16.9\% MRR improvement compared to straightforward Code-Aware Chunking).

Integrating these retrieval configurations into a RAG framework (i.e., when we use an LLM to rerank the dense retrieval candidates) leads to even greater gains, with MRR improving by up to 53\% across chunking strategies. However, the best results are achieved with path-embedded, code-aware chunking. This suggests that while dense retrieval alone may struggle to surface the faulty file in the top positions (e.g., Top-1 or Top-3), ensuring that the correct file appears within the Top-10 or Top-15 candidates allows the LLM to leverage its contextual reasoning capabilities to rerank the files more effectively. \rev{However, for larger values of k (e.g., Top-10), direct dense retrieval with code-aware chunking and file path augmentation outperforms LLM reranking across all chunking strategies.}

\begin{observation}{obs:chunking-performance}
AST-based code-aware chunking outperforms standard text-based chunking for source code. Notably, incorporating relative file paths into the chunks further enhances dense retrieval performance, yielding up to a 20.4\% improvement over traditional text chunking.
\end{observation}

\begin{observation}{obs:rag-improves}
Dense retrieval alone is not sufficient for precise file-level localization. With a complete RAG pipeline, localization accuracy improves substantially—yielding up to a 53\% increase in MRR over standalone dense retrieval.
\end{observation}

\subsubsection{Discussion.} While LLM reranking in the RAG pipeline substantially improves ranking accuracy compared to direct dense retrieval, the effectiveness of the underlying dense retrieval remains a critical factor. It determines whether the relevant file appears within the candidate list at all—since an LLM or agent can only rerank files that have already been retrieved. We observe that path-augmented code-aware chunking yields the most effective retrieval results. For instance, this direct retrieval technique achieves a Top-10 accuracy of 87.3\%, compared to only 70.3\% for standard code-aware chunking—a 24\% improvement. This suggests that incorporating hierarchical path information helps embeddings better capture repository structure and contextual dependencies, improving semantic recall at the file level that is consistent with the design rationale in Section~\ref{sec:repo-encoding}.

\begin{figure}[htbp]
\centering
\includegraphics[width=0.7\linewidth]{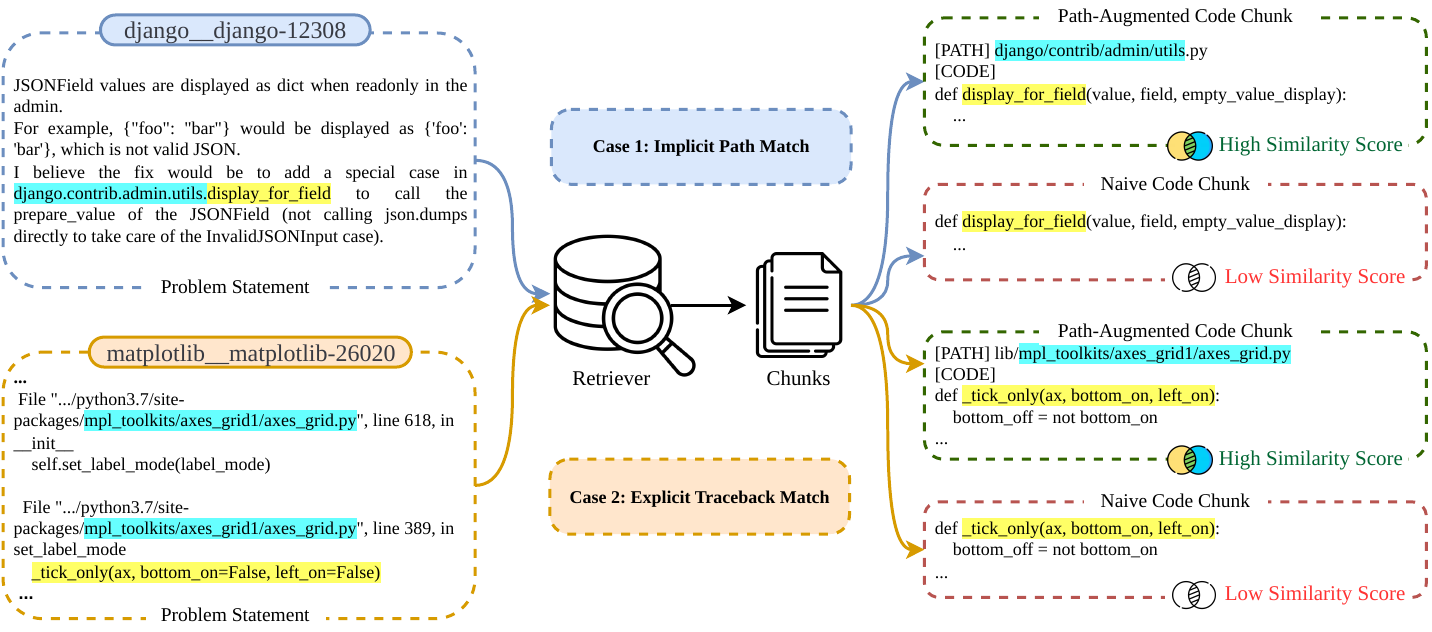}
\caption{Two cases illustrating how path-augmented code chunking improves retrieval 
similarity.}
\label{fig:path-augmented-example}
\end{figure}

\noindent\textbf{\rev{Impact of Path Augmentation.}} \rev{We further analyze how path augmentation improves localization by examining cases where path-augmented chunks correctly surface the faulty file in the top-10 while naive code-aware chunks fail. We identify two prevalent patterns by which bug reports carry path-like signals that augmented chunks can exploit: \textit{implicit path matches}, where package-qualified identifiers in the bug report structurally mirror the repository path hierarchy, and \textit{explicit traceback matches}, where stack trace entries directly reference the file path. In both cases, naive chunks---containing only the function body---lack these signals and rank substantially lower, while path-augmented chunks align with query tokens and achieve higher similarity scores (Figure~\ref{fig:path-augmented-example}).}

\rev{\textbf{\textit{Case 1 (Implicit Path Match).}} In \textit{django\_\_django-12308}, the bug report references \textit{django.contrib.admin.utils}. Although not a verbatim file path, the dot-separated package hierarchy structurally mirrors the repository path \textit{django/contrib/admin/utils.py}. A path-augmented chunk prepending this path aligns with these tokens in the query embedding, yielding a high similarity score and surfacing the correct file. The naive chunk, containing only the function body, lacks this signal and ranks substantially lower.}

\rev{\textbf{\textit{Case 2 (Explicit Traceback Match).}} In \textit{matplotlib\_\_matplotlib-26020}, the bug report contains a direct stack trace entry referencing \textit{mpl\_toolkits/axes\_grid1/axes\_grid.py}. The path-augmented chunk prepended with the relative path of the code matches these tokens directly, producing a high similarity score. The naive chunk, containing only the \textit{\_tick\_only} function body, misses this signal entirely and ranks lower.}


\noindent\textbf{RAG vs.\ Direct Retrieval.} While the traditional RAG pipeline performs well in Top-\{$1-3$\} accuracy, its performance declines for larger $k$. \rev{Specifically, for code-aware chunking with file paths, LLM-based reranking achieves a Top-10 accuracy of 82.3\%, lower than the 87.3\% achieved by direct dense retrieval alone (Table~\ref{tab:retrieval_accuracy})}. We find that the LLM often returns fewer than ten ranked files despite explicit prompt instructions to return 10 files. This behavior likely arises from the model’s tendency to prioritize high-confidence predictions and truncate low-confidence candidates to conserve context length, leading to incomplete Top-$k$ outputs. Overall, while a basic RAG pipeline may significantly improve Top-1 or 3 positions, it may not be as useful for later positions.

\begin{observation}{obs:retrieval-determines}
Although RAG pipelines enhance retrieval accuracy, the effectiveness of dense retrieval still remains a key determinant of overall performance.
\end{observation}



\subsection{RQ1.3 Can agentic reranking improve over traditional RAG?}
\subsubsection{Approach.} In this RQ, we explore whether adding an agent-based reranker can improve over the traditional RAG pipeline. To do this, we adopt the best chunking method ($\star$) identified in \ref{sec:effect-of-retrieval} and compare four setups under the same configuration--(1) a baseline setup without any reranking, where the dense retriever’s initial ranking is used directly; (2) a traditional RAG pipeline that applies LLM-based contextual reasoning over the top-ranked files to select the most relevant one; \rev{(3) \textsc{BLAgent} with agentic scoring only (Phase 1), which uses a ReAct agent to iteratively inspect file skeletons and assign relevance scores; and (4) the full \textsc{BLAgent} pipeline (Phase 1 + Phase 2), which further applies evidence-anchored reranking.}

\begin{table}[htbp]
\centering
\caption{Impact of agentic reranking compared to a traditional RAG pipeline using the base bug report as the retrieval query.}
\label{tab:agentic-reranking}
\small
\begin{tabular}{lcccccc}
\toprule
\textbf{Method} & \textbf{MRR} & \textbf{Top-1} & \textbf{Top-3} & \textbf{Top-5} & \textbf{Top-10} \\
\midrule
No Reranking & 0.553 & 0.417 & 0.627 & 0.713 & 0.873 \\
\midrule
RAG \rev{(LLM Reranking)} & 0.734 & 0.673 & 0.796 & 0.820 & 0.823 \\
\midrule
\textsc{BLAgent} (Phase 1. SAS) & 0.769 & 0.685 & 0.839 & 0.886 & 0.896  \\
\rev{\textsc{BLAgent} (+Phase 2. EAR)} & \textbf{\rev{0.819}} & \textbf{\rev{0.762}}& \textbf{\rev{0.889}} & \textbf{\rev{0.889}} & \textbf{\rev{0.899}}  \\
\bottomrule
\end{tabular}
\end{table}

\subsubsection{Results.} Table~\ref{tab:agentic-reranking} summarizes the comparative results. The traditional RAG pipeline already improves substantially over direct dense retrieval (MRR: 0.734 vs.\ 0.553), confirming that contextual reasoning over retrieved code snippets enhances relevance estimation. However, introducing agentic reranking yields further improvements across all metrics—raising MRR by 11.5\%. This gain indicates that the agentic reranking can better identify and promote truly relevant files that would otherwise remain buried in the candidate list.

\subsubsection{Discussion.} 
\label{sec:rq1-3-discussion}
The results demonstrate that each phase of the reranking pipeline contributes meaningfully and for distinct reasons. Basic RAG with LLM reranking already captures substantial gains through contextual reasoning over retrieved candidates. \rev{Skeleton-based Agentic Scoring (Phase 1) extends this through iterative, hypothesis-driven skeleton inspection, but produces marginal Top-1 gains (+1.8\% over Basic RAG) when structural signatures alone are insufficient to distinguish files that implement similar interfaces or inherit from a common base class. Evidence-anchored reranking (Phase 2) further improves accuracy by constructing pruned file contexts that expand only the method bodies the retriever considers most relevant, grounding the final ranking in implementation-level evidence. The substantial Top-1 improvement from Phase~2 (+11.2\% over Phase~1) confirms that richer, targeted context yields better ranking decisions. Together, the two phases deliver a +11.5\% MRR gain over Basic RAG, with dense retrieval establishing the candidate pool, agentic scoring narrowing it through structural reasoning, and evidence-anchored reranking refining the final ordering with implementation-level precision.}



\begin{observation}{obs:reranking-performance}
Reranking significantly improves localization accuracy over dense retrieval. Agentic reranking further enhances performance through iterative reasoning and structural inspection, while the additional generation-based consolidation stage enables holistic comparison under an expanded but controlled context. Together, these stages yield more consistent Top-$k$ gains and more precise identification of faulty files.
\end{observation}

\subsection{RQ1.4 Can query transformation further improve localization effectiveness?}
\label{sec:rq1-4}
\subsubsection{Approach.} In Section~\ref{sec:effect-of-retrieval}, we discussed that the overall performance of RAG pipelines heavily depends on the quality of dense retrieval. While chunking strategies play a crucial role in preserving code semantics, another key factor influencing retrieval effectiveness is the formulation of the input query used to search the vector database. We investigate whether systematic transformations of bug reports can lead to more effective retrieval. To this end, we experiment with two distinct query transformation strategies (see Section \ref{sec:query-transformation-method}). We apply both transformations on a bug report and then use the transformed reports (i.e., query) as input queries for dense retrieval. The retrieved candidates are then passed to the \textsc{BLAgent} pipeline to evaluate their impact on localization accuracy across both setups.

\subsubsection{Results.} 
\begin{table}[htbp]
\centering
\caption{Impact of query transformation in different settings.}
\small
\label{tab:query-transform}
\begin{tabular}{llcccccc}
\toprule
\textbf{Method} & \textbf{Transformation Type} & \textbf{MRR} & \textbf{Top-1} & \textbf{Top-3} & \textbf{Top-5} & \textbf{Top-10} \\
\midrule
\multirow{3}{*}{Dense Retrieval} & None (Base Bug Report) & 0.553 & 0.417 & 0.627 & 0.713 & 0.873 \\
& Structural ($T_{0}$) & 0.680 & 0.550 & 0.773 & 0.860 & 0.920 \\
& Behavioral ($T_{1}$) & 0.618 & 0.480 & 0.720 & 0.797 & 0.887 \\
\midrule
\multirow{4}{*}{\textsc{BLAgent} (Phase 1)} & None (Base Bug Report) & 0.769 & 0.685 & 0.839 & 0.886 & 0.896  \\
& Structural ($T_{0}$) & 0.785 & 0.696 & 0.856 & 0.903 & 0.930 \\
& Behavioral ($T_{1}$) & 0.783 & 0.696 & 0.860 & 0.900 & 0.910 \\
& \textbf{Both $(T_0 \oplus T_1)$} & \textbf{0.795} & \textbf{0.710} & \textbf{0.860} & \textbf{0.903} & \textbf{0.943} \\
\bottomrule
\end{tabular}
\end{table}

Table~\ref{tab:query-transform} summarizes the impact of query transformation on localization accuracy across both dense retrieval and \textsc{BLAgent}. In the dense retrieval setup, both transformation strategies substantially improve performance over using the base bug report as query (i.e., No Transformation), indicating that enriching the input with syntactic or behavioral cues enhances embedding alignment with relevant code fragments. Among the two, the syntactic transformation ($T_0$) proves more effective (22.9\% MRR improvement compared to using base bug report as query) during direct dense retrieval, suggesting that structural elements—such as module and function identifiers—provide stronger retrieval signals than behavioral descriptions alone.

When these transformations are used within \textsc{BLAgent}, smaller but consistent improvement is realized. Once a relevant file is retrieved within the candidate pool (e.g., Top-15), the agent can typically recover it through reasoning, regardless of its initial rank. Nevertheless, combining both transformations ($T_0 + T_1$) enables the agent to reason over candidates retrieved by complementary signals resulting in a consistent gain. These transformations are particularly valuable when using smaller-context LLMs or when faster reasoning is desired by reducing the number of candidates passed to the model. In such settings, the number of candidates passed to the model is reduced, making it more dependent on the quality of the initial query to ensure that relevant files are retrieved.

\begin{table}[htbp]
\centering
\caption{Sensitivity of BLAgent to candidate ordering retrieved by 
different query transformations.}
\small
\label{tab:query-transform-order}
\begin{tabular}{lcccccc}
\toprule
\textbf{Transformation Order} & \textbf{MRR} & \textbf{Top-1} & \textbf{Top-3} & \textbf{Top-5} & \textbf{Top-10} \\
\midrule
Structural First ($T_0 \oplus T_1$) & 0.795 & 0.710 & 0.860 & 0.903 & 0.943 \\
Behavioral First ($T_1 \oplus T_0$) & 0.797 & 0.716 & 0.866 & 0.903 & 0.930 \\
\bottomrule
\end{tabular}
\end{table}

\noindent\textbf{\rev{Sensitivity to Candidate Ordering.}}
\rev{To assess whether the concatenation order of $T_0$ and $T_1$ introduces positional bias when combining candidates from both transformations in the agentic scoring stage, we evaluate two orderings: (1) $T_0$-first (default), where candidates retrieved by the structural query are prepended before behavioral-query candidates, and (2) $T_1$-first, where this order is reversed. Table~\ref{tab:query-transform-order} shows that both orderings produce nearly identical results across all metrics, with MRR differing by only 0.2\% (0.795 vs.\ 0.797) and Top-1 by 0.8\% (0.710 vs.\ 0.716). This robustness follows from two properties of the pipeline. First, the candidate pool is constructed by appending only unique files from the second transformation that are not already retrieved by the first --- so both orderings yield largely overlapping candidate sets that differ only in which files occupy the marginal positions. Second, the ReAct agent selectively decides which files to inspect based on its evolving reasoning state, rather than processing the list sequentially. Together, these properties ensure that the $T_0$-first default is an implementation convenience rather than a structural dependency of the pipeline.}

\begin{observation}{obs:query-transform}
Query transformations, especially syntactic ($T_0$), significantly boost dense retrieval (22.9\% MRR gain). Using both transformations to select candidates for agentic reranking in the \textsc{BLAgent} pipeline further improves performance across all Top-$k$ ranks.
\end{observation}

\begin{figure}[htbp]
\centering
\includegraphics[width=0.8\linewidth]{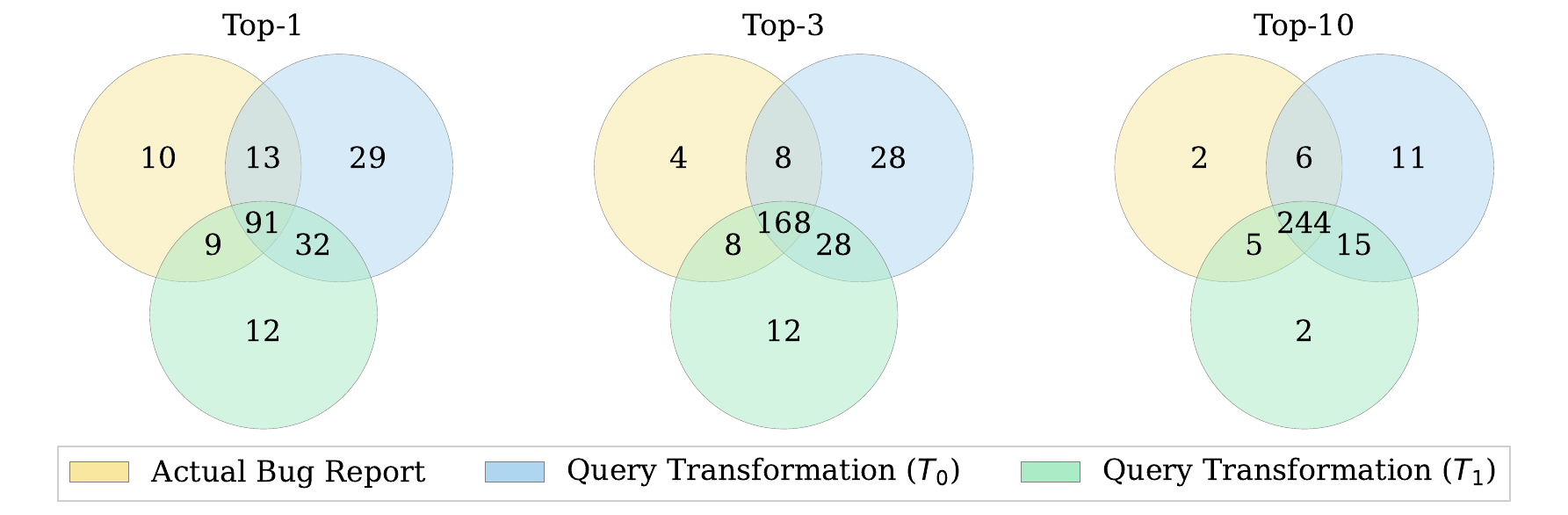}
\caption{File-level localization in dense retrieval when the correct file appears in the Top-{1,3,10} locations.}
\label{fig:file-localization-venn}
\end{figure}

\subsubsection{Discussion.} We further analyze how different transformations impact dense retrieval by examining the overlap and exclusivity of cases where the correct file is ranked highly. Figure~\ref{fig:file-localization-venn} illustrates the complementary nature of the three approaches: the actual bug report, structural query transformation ($T_0$), and behavioral query transformation ($T_1$).

In the Top-1 scenario, both query transformations identify numerous faulty files that the base bug report alone fails to find. Specifically, $T_0$ and $T_1$ uniquely contribute to 29 and 12 Top-1 localizations, respectively, that were not captured by the base report. Additionally, 73 instances are covered by at least one transformed query but missed by the base, demonstrating that transformations are not merely overlapping but in many cases essential for recall. When the window is extended to Top-10, while most of the cases are covered by at least one transformation, there are still 28 cases combined where only query-transformed inputs succeed. However, behavioral transformation becomes less effective on such ranks.

\begin{figure}[htbp]
\centering
\includegraphics[width=0.7\linewidth]{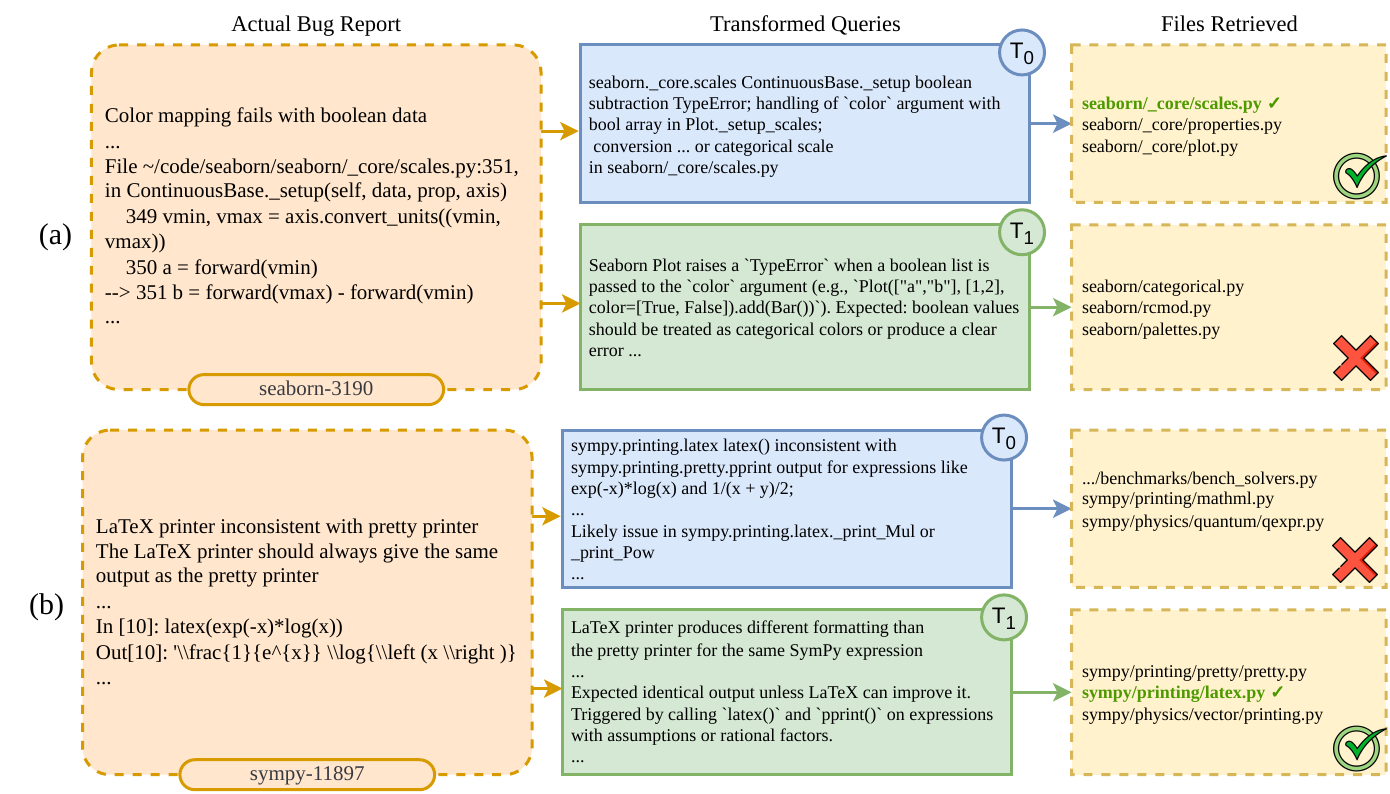}
\caption{Example of different query transformations and retrieved files.}
\label{fig:query-tansformation-example}
\end{figure}

\noindent\textbf{\rev{Complementarity.}} \rev{Across 300 instances, structural queries ($T_0$) retrieve the correct file within Top-10 in 92.0\% of cases and behavioral queries ($T_1$) in 88.7\%, but their \emph{union} achieves 94.3\%---a gain neither type reaches alone. In 17 cases, only $T_0$ retrieves the correct file; in 7 cases, only $T_1$ does. This improvement results from a systematic divergence in retrieval behavior: $T_0$ anchors the candidate set to implementation files by matching on identifiers, traceback tokens, and module references, while $T_1$ broadens it toward observable entry points by aligning with symptom descriptions and runtime behavior. Furthermore, 57.7\% of instances exhibit high complementarity, with at least two unique files in each query's Top-5, providing the downstream agent with a broader and more diverse candidate pool. Figure~\ref{fig:query-tansformation-example} illustrates this complementarity with two representative cases.}

\noindent\textbf{\rev{When Structural Queries Outperform Behavioral Ones.}}
\rev{Structural queries are more effective when the bug report contains an explicit class, 
method, or traceback references that uniquely identify the implementation site. In 
\textit{seaborn-3190} (Figure~\ref{fig:query-tansformation-example} (a), patch: 
\textit{seaborn/\_core/scales.py}), the structural query encodes the precise 
traceback location (\textit{ContinuousBase.\_setup}) and the specific file path, 
retrieving the correct file at Rank~1. The behavioral query, despite naming the 
correct module, disperses similarity mass across color-handling files 
(e.g., \textit{categorical.py}, \textit{palettes.py}) due to 
symptom-level tokens, pushing \textit{scales.py} outside the Top-10. The structural 
query wins because low-frequency, high-specificity tokens from the traceback 
uniquely ground the embedding toward the implementation module.}

\noindent\textbf{\rev{When Behavioral Queries Outperform Structural Ones.}} \rev{Behavioral queries work best in cases where structural cues are weak or misleading---specifically, when the bug manifests as a runtime output mismatch rather than a named API failure. In \textit{sympy-11897} (Figure \ref{fig:query-tansformation-example} (b), patch: \textit{sympy/printing/latex.py}), the structural query targets \textit{\_print\_Mul} and \textit{\_print\_Pow}, focusing on the mathematical equations provided in the bug report and retrieves files responsible for such operations (e.g., \textit{bench\_solvers.py}) which are actually unrelated to this issue. The behavioral query, describing the observed output discrepancy between \texttt{latex()} and \texttt{pprint()} for the same expression, retrieves the correct file at Rank~2 alongside closely related printing modules. The behavioral framing succeeds because it aligns with the module's functional responsibility (output formatting) 
rather than its internal method names.}

\begin{observation}{obs:query-transform-2}
Query transformations provide complementary retrieval signals, substantially enhancing recall over the base bug report. Many faulty files are uniquely retrieved by transformed queries (29 by $T_0$, 12 by $T_1$ in Top-1 out of 300), demonstrating that transformations can be useful for dense retrieval.
\end{observation}



\subsection{RQ1.5 Does the choice of LLM variant affect the performance of the agent?}
\label{sec:llm-variation}
\subsubsection{Approach.}
\rev{To evaluate the robustness of \textsc{BLAgent} across different LLM backbones, we integrate the agentic framework with multiple large language models spanning open-source and closed-source families: \textit{Qwen3-32B}, 
\textit{GPT-OSS-120B}, and \textit{Claude-4.6-Sonnet}}. All models operate on the same query-transformed retrieval process and reasoning protocol with the same temperature (0.7), allowing us to isolate the effect of model scale and architecture on localization performance. We assess file-level accuracy on the SWE-bench-Lite dataset and compare metrics such as MRR and Top-k accuracy to determine whether model scale and type---open-source versus closed-source---provide meaningful gains in the context of agentic reasoning.

\begin{table}[htbp]
\centering
\caption{File-level localization accuracy of \textsc{BLAgent} when equipped with different LLM backbones.}
\label{tab:llm-variants}
\small
\begin{tabular}{llcccccc}
\toprule
\textbf{Dataset} & \textbf{LLM} & \textbf{\#Parameters} & \textbf{MRR} & \textbf{Top-1} & \textbf{Top-3} & \textbf{Top-5} & \textbf{Top-10} \\
\midrule
\multirow{6}{*}{\textbf{SWE-bench-Lite}} 
 & Qwen3 (Phase 1. SAS) & 32B & 0.789 &	0.706	& 0.863 &	0.900 &	0.936 \\
 & Qwen3 (+Phase 2. EAR) & 32B & 0.847 &	0.797	& 0.887 &	0.903 &	0.936 \\
 \cmidrule{2-8}
 & GPT-OSS (Phase 1) & 120B & 0.794	& 0.710	& 0.860	& 0.903 &	0.943 \\
 & GPT-OSS (+Phase 2) & 120B & 0.851 & 0.786 & 0.923 & 0.933 & 0.943 \\
 \cmidrule{2-8}
 & Claude-4.6 (Phase 1) & -- & 0.889 & 0.843 & 0.933 & 0.947 & 0.953 \\
 & Claude-4.6 (+Phase 2) & -- & 0.900 & 0.867 & 0.930 & 0.947 & 0.953 \\
\bottomrule
\end{tabular}
\end{table}

\subsubsection{Results.} Table~\ref{tab:llm-variants} reports localization performance across all three LLM backbones. Among open-source models, despite GPT-OSS-120B being nearly $3.8\times$ larger than Qwen3-32B, Phase~1 MRR differs by only 0.6\% (0.794 vs.\ 0.789), and Top-$k$ accuracies are nearly identical across all thresholds. This indicates that model scale provides only marginal gains within the agentic reasoning framework when operating at the skeleton-inspection level.

\rev{The addition of Phase~2 (evidence-anchored reranking) yields consistent improvements across all three models. Notably, Qwen3-32B benefits most from Phase~2, with MRR improving by 7.3\% and Top-1 by 12.8\% (0.706 $\rightarrow$ 0.797). GPT-OSS-120B shows a similar pattern with a 7.1\% MRR gain. However, the performance for both models in both stages remains almost identical. In contrast, Claude-4.6-Sonnet already achieves strong Phase~1 performance (MRR 0.889, Top-1 0.843), and Phase~2 provides a marginal improvement. This model-dependent pattern suggests that larger, more capable models produce better-separated relevance scores in Phase~1, reducing the marginal contribution of implementation-level evidence in Phase~2, while smaller open-source models rely more heavily on Phase~2 to resolve ambiguous rankings.}

\subsubsection{Discussion.}

These findings reveal a consistent two-tier pattern across model families. For open-source models, agentic reasoning and retrieval coordination are the dominant performance drivers --- the LLM's role is primarily to follow structured reasoning protocols over retrieved skeletal evidence rather than to draw on parametric knowledge, which explains why Qwen3-32B and GPT-OSS-120B perform near-identically in Phase~1 despite a $3.8\times$ size difference. Phase~2 then contributes meaningfully for both, confirming that implementation-level context bridges the gap that skeleton-level scoring leaves open.

\rev{For closed-source frontier models, Claude-4.6-Sonnet demonstrates that stronger reasoning capability translates into more precise skeleton-based scoring, achieving Top-1 84.3\% in Phase~1 alone --- substantially above both open-source models. Phase~2 still improves MRR and Top-1 accuracy, but the gain is smaller, implying that well-separated Phase~1 scores leave less room for consolidation to refine. Practically, this presents a clear deployment trade-off: open-source models with Phase~2 achieve competitive accuracy at lower cost, while frontier models with Phase~1 alone may suffice for a cost-effective solution.}

\begin{observation}{obs:llm-variation}
\textsc{BLAgent} maintains high and consistent performance across both open-source and proprietary LLM backbones. For open-source models, medium-sized models achieve near-equivalent accuracy to larger models, demonstrating that reasoning and retrieval coordination, rather than model scale, are the primary determinants of localization success. \rev{For frontier models, Phase~1 alone achieves strong performance, with Phase~2 providing additional but diminishing improvements.}
\end{observation}

\subsection{RQ1.6 Can BLAgent be extended to function-level localization?}
\label{sec:function-level}

\begin{figure}[htbp]
    \centering
    \begin{minipage}{\linewidth}  
        \begin{lstlisting}[style=prompt, caption={Prompt for function-level localization in BLAgent.}, label={lst:function_loc_prompt}]
The following are the top {len(file_paths)} ranked files retrieved for a given bug report.
Your job is to analyze the python files and method to rerank them based on their relevance to the problem statement.
The idea is to find the actual python method where a patch needs to be applied to fix the bug.

You should return a ranked list of file paths and Class::method names, ordered from most relevant to least relevant,
based on their content and relevance to the problem statement. Think and analyze properly to return the Class::function/method name.
If there is no class, just return the function name. If there is no function, just return the class name.
If there is both, return in Class::method format.
Do not return more than 3 methods per file. You should return at least 10 Class/method names across all files.
Example Output:
```json
{{
    "ranked_files": [
        {{"path/to/most_relevant_file.py": ["Class::most_relevant_method", "second_most_relevant_method"]}},
        {{"path/to/second_most_relevant_file.py": ["second_most_relevant_method"]}},
        ...
        {{"path/to/least_relevant_file.py": ["least_relevant_method"]}}
    ]
}}
```
Do not include any explanations or additional text outside the JSON structure.

Problem Statement:
{problem_statement}
Possibly Relevant Files:
{aggregated_code_text}
"""
\end{lstlisting}
    \end{minipage}
\end{figure}

\subsubsection{Approach.}

\rev{While \textsc{BLAgent} is primarily designed around file-level localization motivated by the requirements of LLM-based APR pipelines, its core components, such as query transformation, AST-aware retrieval, and agentic reranking, are not inherently file-level specific. To evaluate whether the architecture generalizes, we extend \textsc{BLAgent} to function-level localization with a \emph{single modification}: we replace the final consolidated reranking prompt (Listing \ref{lst:reranking_prompt_phase2}) in Phase 2 (EAR) of our agentic reranking stage (Section \ref{sec:agentic-reranking-design}) to instruct the LLM to identify relevant \texttt{Class::method} entities rather than files (Listing~\ref{lst:function_loc_prompt}), keeping all upstream components unchanged. We report Top-$k$ accuracy following prior work~\cite{chang2025bridging, jiang2025cosil}, where a prediction is counted as correct if any predicted method name appears within the method part of any ground-truth \texttt{file::method} string in the Top-$k$ predictions.} \revm{To support a same-model comparison with the strongest baseline, we additionally evaluate \textsc{BLAgent} under Claude-4.6 and compare with LocAgent under the same LLM, mirroring the controlled comparison setup used in Section \ref{sec:agentic-vs-existing}.}

\subsubsection{Results.}

\revm{\textsc{BLAgent} generalizes effectively to function-level localization with a single prompt change, achieving state-of-the-art accuracy under both open-source and frontier LLM settings (Table~\ref{tab:function-level-acc}). With the open-source GPT-OSS model, \textsc{BLAgent} is already competitive with the best prior system, LocAgent (Claude-3.5). Under the same-model comparison, \textsc{BLAgent} (Claude-4.6) establishes a new state of the art across all Top-$k$ thresholds, with the largest margin at Top-1, where it improves over LocAgent (Claude-4.6) by 88\%. Interestingly, LocAgent (Claude-4.6) remains competitive at Top-5 and Top-10 but suffers a sharp Top-1 drop relative to its Claude-3.5 configuration, suggesting that the stronger model surfaces the correct function within its candidate list but ranks it lower within the LocAgent framework. This could be due to the training of the models, of which we cannot speculate. \textsc{BLAgent}, in contrast, improves consistently across all thresholds when moving to Claude-4.6, similar to the trend observed at the file level (Section \ref{sec:agentic-vs-existing}). Hence, BLAgent offers a more stable (besides improved) performance over LocAgent across the models.}

\rev{Importantly, this result is achieved through the single-prompt modification described in Section~\ref{sec:function-level}---all upstream pipeline components remain unchanged---meaning the function-level extension introduces minimal architectural overhead beyond the file-level pipeline. The EAR consolidation step averages only $\approx$15,300 tokens per instance in a single inference pass, similar to the cost reported for file-level localization in \textsc{BLAgent} (Section \ref{sec:cost-analysis}).} \rev{\textsc{BLAgent}'s ability to use an open-source model without extensive fine-tuning further removes dependence on proprietary APIs entirely, making the function-level extension practical for deployment at scale.}

\begin{observation}{obs:func-competitive}
\textsc{BLAgent} achieves state-of-the-art function-level localization performance with a single prompt change and at a fraction of the cost of purpose-built systems, demonstrating that its architecture generalizes beyond file-level without structural modification.
\end{observation}

\begin{table}[htbp]
\centering
\caption{Function-level localization accuracy of \textsc{BLAgent} compared to other approaches on SWE-bench-Lite. $^\dagger$ indicates the approaches we reproduced.}
\label{tab:function-level-acc}
\small
\begin{tabular}{lccc}
\toprule
\textbf{Method} &  \textbf{Top-1} & \textbf{Top-5} & \textbf{Top-10} \\
\midrule
Agentless (GPT-4o) & 0.427 & 0.671 & 0.700 \\
BugCerberus & 0.406 & 0.569 & 0.624 \\
CoSIL (Qwen2.5-32B-FT) & 0.430 & 0.580 & -- \\
OpenHands (Claude-3.5) & -- & 0.682 & 0.700 \\
LocAgent (Claude-3.5) & 0.554 & 0.784 & 0.832 \\
LocAgent (Claude-4.6)$^\dagger$ & 0.386 & 0.765 & 0.841 \\
\midrule
\textsc{BLAgent} (GPT-OSS) & 0.536 & 0.781 & 0.810 \\
\textsc{BLAgent} (Claude-4.6) & \textbf{0.726} & \textbf{0.861} & \textbf{0.872} \\
\bottomrule
\end{tabular}
\end{table}

\subsubsection{Discussion.}

\rev{The strong function-level result is best understood as a direct consequence of BLAgent's file-level precision rather than the prompt modification alone. BLAgent retrieves the correct patch file within its Top-5 candidates in over 90\% of all instances, and once the correct file is present, the problem of identifying relevant methods becomes a tightly constrained search. When the candidate file set is accurate and compact, function-level---and by extension statement-level---prediction tasks benefit from a reduced search space, lower context pressure on the LLM, and a cleaner signal-to-noise ratio in the retrieved evidence.} \revm{The further gains from GPT-OSS to Claude-4.6 (35.4\% improvement at Top-1) reinforce that, given a high-precision candidate set, function-level accuracy scales primarily with the underlying LLM's reasoning capability rather than with additional architectural complexity.}

\rev{This points to a natural integration opportunity. Rather than treating file-level and function-level localization as competing tasks, a cascaded design---in which BLAgent's ranked files serve as input to a graph-guided function-level agent such as LocAgent or Agentless---would allow each component to operate in its strongest regime: BLAgent contributes high-precision, low-cost file discovery, while the downstream agent focuses its traversal within a pre-filtered, high-confidence candidate set. Such a cascade could outperform either system in isolation while remaining cost-efficient, and is a promising direction for future hierarchical repair pipelines.}

\begin{observation}{obs:func-filelevel-foundation}
Precise file-level localization directly enables fine-grained localization by tightly constraining the function search space.
\end{observation}



\section{Impact Assessment of BLAgent Bug Location on Automated Bug Repair (RQ2)}

To better understand how \textsc{BLAgent} localization influences downstream repair effectiveness, we break down RQ2 into the following sub-RQs:

\begin{itemize}
    \item \textbf{RQ2.1: Does localizing with \textsc{BLAgent} improve program repair?}  
    
    We investigate whether replacing the baseline localization module with \textsc{BLAgent} leads to a higher proportion of correct repairs in Agentless, thereby assessing the causal impact of improved localization on overall APR effectiveness.

    \item \textbf{RQ2.2: At which stage does the repair process fail even with correct localization?}  
    
    We decompose the end-to-end APR pipeline to identify where failures arise—across different levels of localization (e.g., file, function, line) and patch generation—and analyze why these issues occur despite accurately identifying the faulty file.
    
\end{itemize}

\subsection{Experimental Setup}
\subsubsection{Agentless APR Framework.} We adopt Agentless~\cite{xia2025agentless} as our downstream APR baseline due to several reasons. First, it is widely used both in the industry (e.g., OpenAI) and academia \cite{chen2025locagent, chang2025bridging, jiang2025cosil} to create and compare bug repair patches. Second, it is highly modular with a clear separation between localization and repair components. Third, Agentless has demonstrated competitive performance on the SWE-bench dataset, making it a representative baseline for modern LLM-based APR systems. While Agentless itself is non-agentic, we use it purely as a downstream APR framework to evaluate the impact of BLAgent’s enhanced file-level localization. Since the original framework does not natively support Ollama-based LLMs or the embedding models used in our setup, we extend it to interface with the Ollama runtime and the HuggingFace embedding library. All other components of the Agentless pipeline, including patch generation, validation, and evaluation, remain unchanged to ensure comparability. We use the same GPT-OSS-120B model as the base LLM and nomic-embed-text-v1 as the embedding model (similar to Section \ref{sec:paper-baselines}).


\begin{figure}[htbp]
    \centering
    \includegraphics[width=0.7\linewidth]{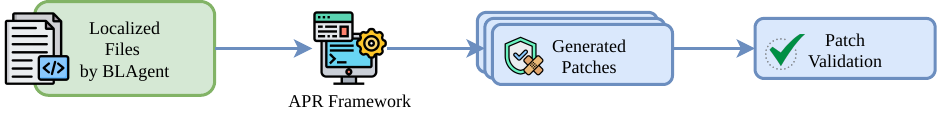}
    \caption{Integration of BLAgent into another APR framework.}
    \label{fig:overall-repair-outline}
\end{figure}

For this evaluation, we first execute Agentless using its native localization module as a baseline. Then, we replace Agentless' native file-localization approach with \textsc{BLAgent}. Given a bug report and a target repository, our framework produces a ranked list of candidate files, assigning each file a relevance score via agentic reranking. The top-3 files (matching Agentless’s default) are passed to the Agentless repair pipeline. From this stage onward, the workflow follows the standard Agentless protocol: line-level localization is performed within the selected files using targeted LLM prompting, after which 40 candidate patches are generated using an LLM through a high temperature of 0.8 (default configuration). The final patch is then selected via majority voting, with regression and reproduction tests employed to validate the correctness of the repair. The overall setup is shown in Figure \ref{fig:overall-repair-outline}.

\subsubsection{Evaluation Metrics}
Following the evaluation protocol of SWE-bench~\cite{jimenez2024swebench}, we assess the effectiveness of program repair using the \textbf{Resolved Rate}. This metric represents the proportion of bug reports for which the generated patches successfully pass all the validation tests. Let $N_{\text{resolved}}$ denote the number of resolved bugs and $N_{\text{total}}$ the total number of bugs. Then,
\begin{equation}
\mathrm{Resolved\ Rate} = \frac{N_{\text{resolved}}}{N_{\text{total}}} \times 100\%.
\end{equation}
A higher resolved rate indicates a stronger end-to-end repair capability of the system.

\subsection{RQ2.1 Does localizing with \textsc{BLAgent} improve program repair?}
\label{sec:rq-repair-accuracy}

\subsubsection{Approach.}
We integrate \textsc{BLAgent} into the Agentless framework~\cite{xia2025agentless} and evaluate it on the SWE-bench-Lite dataset~\cite{jimenez2024swebench}. Similar to the original work, we conduct experiments under three configurations: (i) \emph{Majority Voting}, which generates 40 patch candidates and selects the final one via majority voting; (ii) \emph{+Regression Test}, where available regression tests are included; and (iii) \emph{+Reproduction Test}, which further generates and incorporates reproduction tests to select a generated patch.


\begin{table}[htbp]
\centering
\caption{Resolution rate on SWE-bench-Lite dataset with different methods \rev{(Best result is reported for both methods across 3 runs)}.}
\label{tab:resolution}
\small
\begin{tabular}{lcccccc}
\toprule
\textbf{APR} & \textbf{Patch Selection} & \textbf{Localization Method} & \textbf{\#Instances} & \textbf{\#Resolved} & \textbf{Rate} \\
\midrule 
 \multirow{6}{*}{Agentless} & \multirow{2}{*}{Majority Voting} & Native & 300 & 83 & 27.6\%\\
  & & \textsc{BLAgent} & 300 & 102 ($\uparrow$22.8\%) & 34.0\%\\
 \cmidrule(lr){2-6}
  & \multirow{2}{*}{+Regression Test} & Native & 300 & 86 & 28.6\% \\
  & & \textsc{BLAgent} & 300 & 108 ($\uparrow$25.5\%) & 36.0\% \\
 \cmidrule(lr){2-6}
  & \multirow{2}{*}{+Reproduction Test} & Native & 300 & 96 & 32.0\% \\
  & & \textsc{BLAgent} & 300 & 115 ($\uparrow$19.7\%) & 38.3\% \\
\bottomrule
\end{tabular}
\end{table}

\subsubsection{Results.}
Table~\ref{tab:resolution} summarizes the results. Integrating \textsc{BLAgent} consistently improves the end-to-end repair rate across all settings. \rev{We run the framework with both localization approaches three times and report the best result for both methods in Table \ref{tab:resolution}.} Under the majority voting setup, our approach resolves up to 102 bugs versus 83 by the baseline (comparable to the 79 originally reported by \textsc{Agentless} with GPT-4o), representing a relative improvement of over 20\%. When regression and reproduction tests are included, a similar performance gap persists.

A closer examination of the generated patches reveals another important benefit. When using GPT-OSS-120B as the base LLM, the baseline framework produced 38 empty patches—instances where no valid patch was ultimately selected—compared to \rev{14} when we used the \textsc{BLAgent} as the file-level localization method. Empty patches often indicate that the LLM received insufficient or irrelevant context, either due to inaccurate localization or because the generated patch could not be parsed as a valid fix. The reduction in such cases suggests that the proposed method provides more semantically grounded input, allowing the LLM to reason more effectively about the underlying fault and produce syntactically valid, contextually coherent patches.

We note that a subset of issues (e.g., 23 issues from \texttt{matplotlib}) could not execute regression or reproduction tests due to the reliance of Agentless on an obsolete SWE-bench 2.1 interface, which is now outdated. Attempts to migrate to the latest version revealed compatibility issues requiring nontrivial engineering effort, which we leave for future work. Therefore, the true repair potential is likely higher than reported, as these missing cases could result in additional successful patches.

Although \textsc{BLAgent} achieves a Top-1 file-level localization accuracy close to 80\% with GPT-OSS, the corresponding end-to-end resolution rate remains around 38\%. This disparity highlights a key insight: accurate file-level localization, while necessary, is not sufficient for successful repair. Downstream steps such as line-level localization and semantic patch generation in the APR pipeline remain major limiting factors.

\begin{figure}[htbp]
\centering
\includegraphics[width=0.9\linewidth]{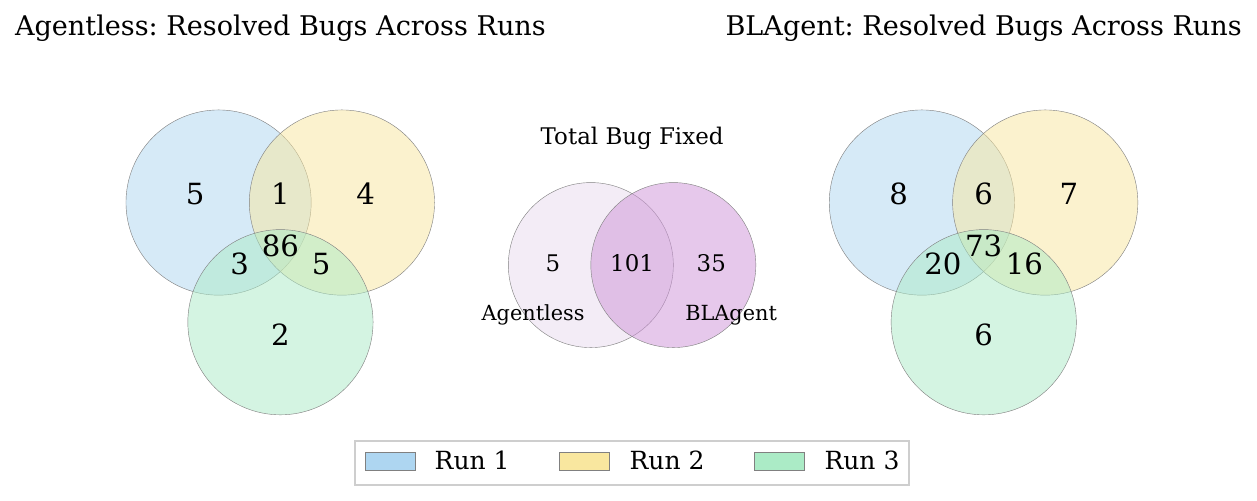}
\caption{Overlap of repaired issues across multiple runs using different localization strategies.}
\label{fig:repair-overlap}
\end{figure}

\begin{observation}{obs:repair-performance}
Integrating \textsc{BLAgent} into \textsc{Agentless} consistently improves end-to-end repair rates (often over 20\% gain) and reduces empty patches. While precise file localization substantially facilitates successful repair, ultimate patch correctness also depends on subsequent stages within an APR.
\end{observation}
\begin{observation}{obs:repair-performance-when-test-integrated}
Integrating regression and reproduction tests for patch selection consistently enhances repair performance, independent of the localization method, by guiding the APR system toward correct and verifiable patches.
\end{observation}

\subsubsection{Discussion.}
While \textsc{BLAgent} consistently improves repair performance, we next investigate the stability of these improvements. Figure~\ref{fig:repair-overlap} compares repair overlaps across multiple runs using both the native and \textsc{BLAgent} localizations. We observe some divergence between runs, with each resolving partially disjoint sets of issues. This instability can largely be attributed to the stochasticity inherent in high-temperature decoding during both patch and test generation, suggesting that overall resolved counts alone may obscure deeper behavioral patterns of APR systems. \rev{To quantify whether the observed improvements are statistically meaningful, we apply a paired Wilcoxon signed-rank test over per-bug solve frequencies across the three runs, confirming that the observed improvement is statistically significant ($p = 0.014$) with $p < 0.05$. Across all 3 runs, using BLAgent as the localization module leads to 136 unique total bug fixes (45.3\%) in the SWE-bench-Lite dataset, where it solves 35 unique bugs, while the native localization method only solved 106 (35.3\%).}

\begin{table}[h]
\centering
\caption{Unique repairs per method and frequency with which the opposite method failed to localize the correct file. Values represent (incorrect localizations on the opposing method / total unique repairs on a method).}
\label{tab:repair_localization_matrix}
\small
\begin{tabular}{lccc}
\toprule
\textbf{Loc Method} & \textbf{Run 1 (Incorrect (I) / Total (T))} & \textbf{Run 2 (I / T)} & \textbf{Run 3 (I / T)} \\
\midrule
BLAgent Loc. (vs. Agentless Loc.) & 12 / 34 (35.3\%) & 11 / 24 (45.8\%) & 17 / 34 (50.0\%)\\
Agentless Loc. (vs. BLAgent Loc.)     & 0 / 22 (0.0\%)  & 1 / 18 (5.6\%)  & 1 / 15 (6.7\%)\\
\bottomrule
\end{tabular}
\end{table}

Furthermore, we compare the results from multiple runs of the Agentless APR framework—once using its native localization method and once using \textsc{BLAgent} for file-level localization. Specifically, we focus on the \textit{uniquely repaired} issues, cases successfully fixed when using one localization method but not the other. For these issues, we examine whether the alternative method failed to correctly identify the faulty file. Table~\ref{tab:repair_localization_matrix} summarizes this comparison by showing how often the other method failed to localize the correct file for these uniquely repaired cases. When our approach is used during localization, and it exclusively repairs an issue, the baseline localization method (Agentless) often fails to correctly localize the faulty file (e.g., 50\% of such cases in Run~3). In contrast, for issues uniquely repaired by Agentless when using its own localization method, \textsc{BLAgent} localized the correct file in nearly all instances. This pattern shows that unique repairs resulting from the BLAgent's localization frequently arise from scenarios where the baseline’s localization was insufficient, suggesting that improved file-level grounding directly enables successful patch synthesis. Conversely, the few cases where Agentless succeeded despite our correct localization indicate failures in later APR stages—such as patch generation—rather than in file-level localization itself.  

Overall, these findings highlight that accurate localization is not a peripheral factor but a foundational determinant of downstream repair success. While improved localization does not guarantee a correct patch, it substantially increases the probability of success by ensuring the LLM operates on the correct fault context. When the faulty file is missing from the retrieved candidates, even advanced reasoning cannot compensate.

\begin{observation}{obs:repair-diff}
Repair outcomes in the APR framework exhibit some stochastic variation across runs, resolving partially disjoint issue sets. This instability shows the sensitivity of LLM-based APRs to sampling randomness in patch and test generation.
\end{observation}

\begin{observation}{obs:repair-and-loc-relation}
Most unique repairs obtained with \textsc{BLAgent} occur when the baseline localization method fails to localize the correct file, demonstrating a direct causal link between accurate localization and successful repair. This finding emphasizes that localization quality is a decisive factor in end-to-end APR effectiveness, not just a secondary component.
\end{observation}

\subsection{RQ2.2 At which stage does the repair process fail even with correct localization?}

\subsubsection{Approach.} To better understand where and how the program repair framework fails despite high file-level localization accuracy, we conduct a detailed error analysis across Agentless’s hierarchical pipeline. Specifically, we aim to identify whether unresolved cases stem from failures in localization (at file, function, or line level) or from the subsequent patch generation step. \rev{For this experiment, we analyze the artifacts from Run 2 (Section~\ref{sec:rq-repair-accuracy}), where \textsc{BLAgent} was used as the localization method and the lowest number of bugs (102) were resolved across the three runs. Focusing on the worst-performing run provides a conservative view of the failure modes, ensuring that the identified bottlenecks are not artifacts of a particularly favorable run.}

We begin by selecting 177 unresolved bug instances—cases where Agentless failed to produce a correct patch—and systematically trace their progression through each stage of the framework. Using Agentless’s default configuration, the model considers the top-$k$ ranked outputs from each previous stage as context for the next (with $k=3$ in our experiments). For example, during function-level localization, the model only explores functions from the top-3 retrieved files; for line- or statement-level localization, it only considers the top-3 functions, and so on. This hierarchical dependency makes error propagation particularly important to diagnose.

For each stage, we compare model predictions with ground-truth locations extracted from the actual developer patches. At the file level, localization is considered correct if the ground-truth file appears among the top-3 retrieved candidates. The same criterion is applied to function-level localization. However, evaluating line-level accuracy is more nuanced, since small semantic differences can lead to valid fixes at slightly different line numbers or equivalent statements. To address this, we employ a lightweight heuristic approximation to estimate line-level localization accuracy.

\noindent\textbf{Line-level Approximation.}
The approximation algorithm compares the first changed statement in the model-generated patch with that in the ground-truth patch. It proceeds as follows:
\begin{enumerate}
\item Extract the first code line (excluding comments and docstrings) that was added or removed in both patches.
\item If both edits occur within the same file and their line numbers differ by at most 5 lines, line-level localization is considered correct.
\item Otherwise, compute the normalized sequence similarity between the two modified lines. We straightforwardly calculate the similarity using \texttt{SequenceMatcher} \footnote{\url{https://docs.python.org/3/library/difflib.html\#difflib.SequenceMatcher}}.
\item If the edits belong to the same file, exhibit textual similarity above a threshold (similarity $\geq 0.6$), and occur within a small positional window (e.g., within 5–10 lines), we treat the result as an approximate match at the line level. \rev{The threshold of 0.6 was determined by manually inspecting a small sample of cases, confirming that it reliably captures genuine near-miss localizations.}
\end{enumerate}

This heuristic allows us to capture cases where the framework localizes the correct logical region but introduces the modification slightly above or below the true location—an issue common in real repair settings. While this approximation does not provide an exact measure of line-level accuracy, it offers a pragmatic way to distinguish between near-miss localizations and genuine mislocalizations. Similar strategies have been adopted in prior studies to approximate line-level localization accuracy \cite{xia2025agentless}.

\begin{figure}[htbp]
\centering
\includegraphics[width=0.8\linewidth]{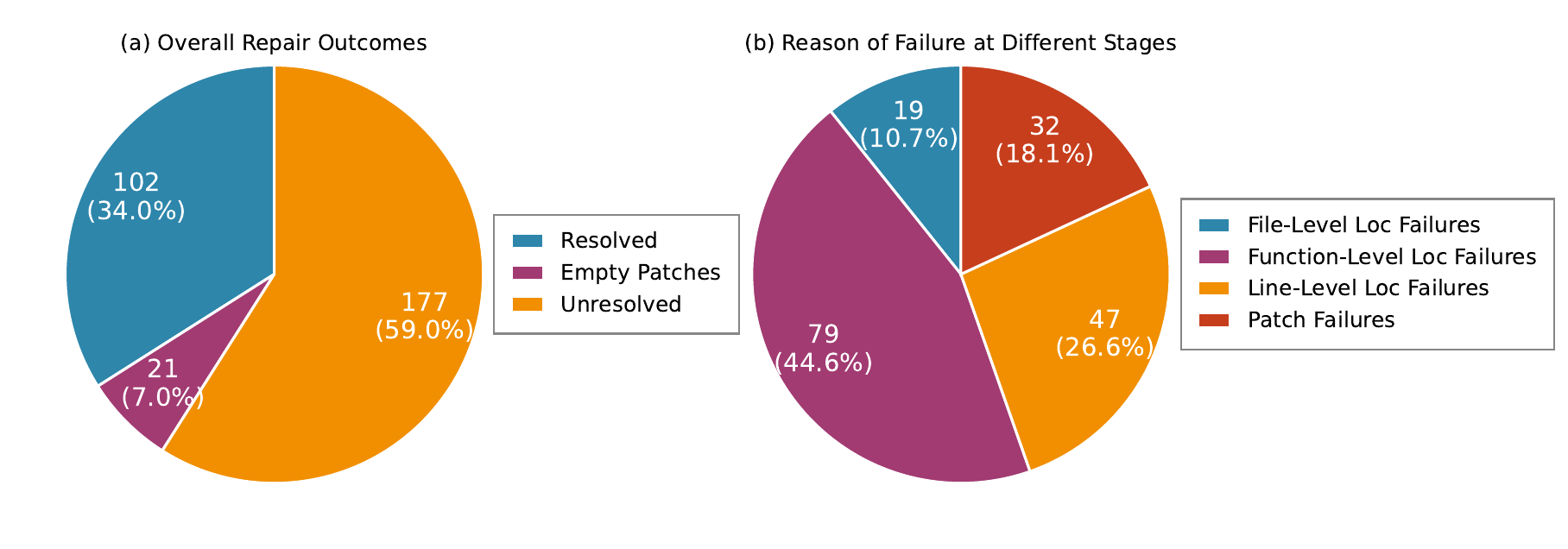}
\caption{Overall resolution and failure percentage at different levels of program repair stage.}
\label{fig:repair-outcomes}
\end{figure}

\subsubsection{Results.}

As shown in Figure ~\ref{fig:repair-outcomes}a, approximately one-third of all instances (34\%) were successfully resolved by the system. However, a majority of the cases remain unresolved, including a smaller fraction (7\%) of empty patches (i.e., no patch was selected at the end), suggesting that the APR framework often fails to produce a syntactically and logically valid modification. The unresolved cases (almost 60\%) where a patch was generated but could not fix the bug motivate a deeper look into the repair pipeline to understand where failures originate. Out of the 177 unresolved cases, we find that 158 issues were localized correctly at the file-level (Figure \ref{fig:repair-outcomes}b) with \textsc{BLAgent} but failed in a later stage in the program repair process.

\noindent\textbf{Failures Across Localization Stages.} We decompose
unresolved cases into four categories in Figure~\ref{fig:repair-outcomes}(b), reflecting the specific stages at which the program repair pipeline fails. Most of these issues occurred in one of the stages of the hierarchical localization pipeline. 

\noindent\textbf{File-level localization failures} constitute 10.7\% of the unresolved cases, indicating that a subset of bug reports still leads the retrieval module toward incorrect source files. These failures predominantly occur when the correct file does not appear among the Top-3 retrieved candidates. While expanding the retrieval depth could mitigate some of these errors, it would also increase the computational overhead and complexity of the Agentless framework, creating a trade-off between coverage and efficiency.

\begin{figure}[htbp]
    \centering
    \begin{minipage}{0.7\linewidth}  
        \begin{lstlisting}[style=prompt, caption={Example of function-level localization failure due to parsing issue.}, label={lst:loc-failure-func-level}]
Expected Format:
sympy/core/expr.py
function: Expr.__eq__

Actual LLM Output:
sympy/core/expr.py: Expr.__eq__
sympy/core/sympify.py: sympify
sympy/core/sympify.py: _sympify
sympy/core/sympify.py: converter

\end{lstlisting}
    \end{minipage}
\end{figure}

In our observation, the most frequent failures occur at the \textbf{function-level localization} (44.6\%), highlighting persistent challenges in accurately ranking the correct function among the Top-3 candidates even when the faulty file itself is correctly retrieved. A notable source of these errors arises from formatting inconsistencies in LLM outputs that prevent the APR framework from correctly parsing the generated function names. For instance, in \texttt{sympy\_\_sympy-18057}, although the correct function name appeared in the LLM’s response, the framework failed to interpret it due to a structural mismatch, leading to no valid function being selected (see Listing~\ref{lst:loc-failure-func-level}).

\begin{figure}[htbp]
    \centering
    \begin{minipage}{0.7\linewidth}  
        \begin{lstlisting}[style=prompt, caption={Example of line-level localization failure.}, label={lst:loc-failure-line-level}]
Localization Log:
2025-10-17 00:58:42,983 - INFO - Creating Ollama API request
2025-10-17 00:59:56,302 - INFO - Ollama API response: ```
src/flask/cli.py
line: 1021
line: 1022
...
line: 1034
```

Ground Truth Patch:
--- a/src/flask/cli.py
+++ b/src/flask/cli.py
@@ -9,7 +9,7 @@
...
@@ -989,49 +989,62 @@ def shell_command() -> None:
@click.option(
     "--sort",
     "-s",
...

\end{lstlisting}
    \end{minipage}
\end{figure}

The second most common failures occur at the \textbf{line level localization} (26.6\%), where the model successfully identifies the faulty function but struggles to locate the precise statement or edit region within it. These near-miss cases indicate that while the model captures the correct semantic region, it lacks sufficient fine-grained reasoning about program behavior and fix placement. The underlying causes of such line-level failures can be diverse—ranging from intrinsic LLM limitations in reasoning about code execution to occasional hallucinations \cite{huang2025survey} or incomplete code understanding when the LLM outputs completely unrelated line numbers. In our investigation, we also observed cases such as \texttt{pallets\_\_flask-5063}, where the ground-truth patch modifies multiple methods, but the LLM focused only on a single one (lines 1021–1034), resulting in a partial and ultimately incorrect localization (see Listing~\ref{lst:loc-failure-line-level}).

Finally, \textbf{patch-generation failures} (18.1\%) occur when localization is successful at all steps, but the code synthesis stage either produces invalid or semantically incorrect edits. This points to the limitations of current generation models in maintaining compilation consistency and applying contextually coherent changes, even with accurate localization cues. Figure~\ref{fig:patch-diff} provides a concrete example from \texttt{sympy\_\_sympy-22005}. Here, the localization is accurate down to the line level, and the generated patch targets the same location as the ground-truth patch. However, the generated patch (Figure~\ref{fig:patch-diff}a) adds a conditional check to restrict univariate polynomials to systems with a single generator. While syntactically correct, it does not fully encode the intended semantics. In contrast, the ground-truth patch (Figure~\ref{fig:patch-diff}b) raises a \texttt{NotImplementedError} for underdetermined systems, enforcing the zero-dimensional constraint and preserving correct program behavior.

\begin{figure}[h!]
\centering
\begin{minipage}[t]{0.49\textwidth}
\centering
\caption*{\textbf{(a) Generated Patch}}
\begin{lstlisting}[style=diff]
diff --git a/sympy/solvers/polysys.py b/sympy/solvers/polysys.py
index 66efb1a..8ab9101 100644
--- a/sympy/solvers/polysys.py
+++ b/sympy/solvers/polysys.py
@@ -240,7 +240,9 @@ def solve_generic(polys, opt):
    univariate = list(filter(_is_univariate, basis))
-   if len(univariate) == 1:
+   # Only accept a single univariate polynomial when the system has exactly one generator.
+   # This prevents treating underdetermined (infinite-solution) systems as zero-dimensional.
+   if len(univariate) == 1 and len(gens) == 1:
        f = univariate.pop()
\end{lstlisting}
\end{minipage}
\hfill
\begin{minipage}[t]{0.49\textwidth}
\centering
\caption*{\textbf{(b) Ground-truth Patch}}
\begin{lstlisting}[style=diff]
diff --git a/sympy/solvers/polysys.py b/sympy/solvers/polysys.py
--- a/sympy/solvers/polysys.py
+++ b/sympy/solvers/polysys.py
@@ -240,6 +240,12 @@ def _solve_reduced_system(system, gens, entry=False):
    univariate = list(filter(_is_univariate, basis))
 
+   if len(basis) < len(gens):
+       raise NotImplementedError(filldedent('''only zero-dimensional systems supported (finite number of solutions)'''))
+
         if len(univariate) == 1:
\end{lstlisting}
\end{minipage}
\caption{Comparison of (a) APR-generated incorrect patch, 
and (b) Ground-truth patch for \texttt{sympy\_\_sympy-22005}.}
\label{fig:patch-diff}
\end{figure}

Overall, our evaluation indicates that while end-to-end APR failures arise at multiple stages, localization constitutes a major bottleneck. File-level, function-level, and line-level localization collectively account for the majority of unresolved cases. Notably, in our analysis, approximately 82\% of unresolved issues could be traced to failures at some stage of hierarchical localization. These findings highlight that precise bug-context localization is a necessary prerequisite for effective LLM-based program repair, and that improvements in both fine-grained localization and code synthesis are essential to substantially increase repair success.

\begin{observation}{obs:hierarchical-failure}
Failures in hierarchical localization are the main source of unresolved APR cases, with function-level (44.6\%) and line-level (26.6\%) errors being the most common. Accurate identification of the faulty file, function, and line is therefore critical to enable successful patch generation.
\end{observation}

\begin{observation}{obs:parsing-failure}
Parsing errors in LLM outputs block correct function selection, propagating failures downstream and limiting APR effectiveness.
\end{observation}

\begin{observation}{obs:patch-failure}
Even when localization is correct, 18.1\% of failures arise from patch-generation errors, where the model produces invalid or semantically incorrect edits. This shows that while precise localization is necessary, the success of APR also depends on the LLM’s ability to reason about program logic and generate contextually correct patches.
\end{observation}

\subsubsection{Discussion.} 
Given that most unresolved cases stem from function- or line-level localization errors, we investigate whether these failures are purely localization-related and if providing correct statement-level context could lead to successful repair. 

We analyze the case of \texttt{django\_\_django-11133}, where file-level localization was correct but line-level localization failed in one of the experimental runs. During the failed run, the APR system incorrectly targeted line 309 (Figure~\ref{fig:line-level-patch-failure}a) instead of the correct region near line 229 (Figure~\ref{fig:line-level-patch-failure}b). As a result, the generated patch was syntactically valid but semantically irrelevant, and the repair failed.

\begin{figure}[h!]
\centering
\begin{minipage}[t]{0.48\textwidth}
\centering
\caption*{\textbf{(a) Patch resulted from inaccurate localization}}
\begin{lstlisting}[style=diff]
--- a/django/http/response.py
+++ b/django/http/response.py
@@ -309,7 +309,8 @@ class HttpResponse(HttpResponseBase):
     @content.setter
     def content(self, value):
\end{lstlisting}
\end{minipage}
\hfill
\begin{minipage}[t]{0.48\textwidth}
\centering
\caption*{\textbf{(b) Ground-truth Patch}}
\begin{lstlisting}[style=diff]
--- a/django/http/response.py
+++ b/django/http/response.py
@@ -229,7 +229,7 @@ def make_bytes(self, value):
        # Handle string types -- we can't rely on force_bytes here...
\end{lstlisting}
\end{minipage}
\caption{Example of failed line-level localization (\texttt{django\_\_django-11133}).}
\label{fig:line-level-patch-failure}
\end{figure}

To test whether precise localization alone could resolve the issue, we re-executed the repair process by replacing the model’s predicted line number with the correct one. With accurate line-level context, the APR successfully generated the correct patch (Figure~\ref{fig:line-level-patch-fix}), demonstrating that the previous failure was not due to deficiencies in patch synthesis but rather to mislocalized edit positions. This case study highlights a key insight: deviations in statement- or line-level localization can derail the repair process entirely.

\begin{figure}[h!]
\centering
\begin{minipage}[t]{\textwidth}
\centering
\begin{lstlisting}[style=diff]
diff --git a/django/http/response.py b/django/http/response.py
index 5126111..fa61f0b 100644
--- a/django/http/response.py
+++ b/django/http/response.py
@@ -226,14 +226,18 @@ class HttpResponseBase:
# [SOME COMMENTS ARE TRUNCATED TO SAVE SPACE]
+        if isinstance(value, (bytes, bytearray, memoryview)):
+            return bytes(value)
+
-        if isinstance(value, bytes):
-            return bytes(value)
         if isinstance(value, str):
             return bytes(value.encode(self.charset))
-        # Handle non-string types.
+
+        # Handle non-string, non-bytes-like types.
         return str(value).encode(self.charset)
 
     # These methods partially implement the file-like object interface.
\end{lstlisting}
\end{minipage}
\caption{Generated patch with correct line-level information.}
\label{fig:line-level-patch-fix}
\end{figure}

\begin{observation}{obs:loc-corrects-patch}
Providing the correct statement-level context allows the APR system to generate a correct patch, demonstrating that accurate localization is a critical enabler of effective reasoning and patch synthesis in LLM-based APR pipelines.
\end{observation}


\section{Impact of Key Design Decisions in BLAgent}
\subsection{Cost Analysis}
\label{sec:cost-analysis}

\rev{Table~\ref{tab:cost-analysis} reports the average token consumption and 
estimated API cost per bug instance in USD for BLAgent across two model backends. For the Claude-4.6-Sonnet model, we calculate the cost using its official pricing documentation\footnote{\url{https://platform.claude.com/docs/en/about-claude/pricing} (Input \$3/M Tokens; Output \$15/M Tokens)}, and for the GPT-OSS-120B model, we estimate the cost using a third-party provider \footnote{\url{https://novita.ai/models/model-detail/openai-gpt-oss-120b} (Input \$0.05/M Tokens; Output \$0.25/M Tokens)}. With GPT-OSS-120B, the full agentic reranking pipeline consumes 24,523 prompt tokens and 1,939 completion tokens per bug, corresponding to an estimated cost of \$0.0017 per instance. With Claude-4.6-Sonnet, the total consumption is 26,180 prompt tokens and 582 completion tokens, at an estimated \$0.09 per instance. Across all 300 SWE-bench-Lite instances, the total localization cost amounts to less than \$1 with GPT-OSS-120B and \$27 with Claude-4.6-Sonnet.}

\begin{table}[htbp]
\centering
\caption{Per instance cost analysis of BLAgent using different LLMs.}
\label{tab:cost-analysis}
\small
\begin{tabular}{llccc}
\toprule
\textbf{Model} & \textbf{Reranking Phase} & \textbf{Prompt Tokens} & \textbf{Completion Tokens} & \textbf{Cost (\$)} \\
\midrule
\multirow{3}{*}{\textbf{GPT-OSS}} 
& Phase 1  & 8,550 & 1,866  & 0.0009 \\
& Phase 2   & 15,973 & 72    & 0.0008 \\
& \textbf{Total} & 24,523	& 1,939 & \textbf{0.0017} \\
\midrule
\multirow{3}{*}{\textbf{Claude-4.6}}
& Phase 1  & 6,453 & 483  & 0.03 \\
& Phase 2   & 19,727 & 99    & 0.06 \\
& \textbf{Total} & 26,180	& 582 & \textbf{0.09} \\
\bottomrule
\end{tabular}
\end{table}

\noindent\textbf{\rev{Comparison with LocAgent.}}
\rev{To contextualize BLAgent's cost efficiency, we ran LocAgent under identical conditions using Claude-4.6-Sonnet with a fixed budget of \$300 for the 300 instances in the SWE-bench-Lite benchmark. LocAgent exhausted the full \$300 budget after processing only 182 instances, producing valid results for 158 of them. The remaining instances either timed out or exceeded 800,000 tokens/minute per instance due to unbounded graph traversal loops before the budget was depleted. Among the 158 completed instances, LocAgent consumed an average of 282,720 prompt tokens and 2,634 completion tokens per bug, at an average cost of \$0.89 per instance. However, this figure understates the true cost, as failed instances consumed disproportionately more tokens---bringing the observed cost across all 182 attempted instances to \$1.65 per instance (\$300 $\div$ 182). In contrast, BLAgent completed all 300 instances at \$0.09 per instance (\$27 total), representing a cost reduction of more than 18$\times$ relative to LocAgent's observed rate.}

\begin{observation}{obs:cost-analysis}
BLAgent's bounded agentic reasoning is substantially more cost-efficient than complex graph-traversal-based approaches: BLAgent completes all 300 benchmark instances at \$0.09 per bug with Claude-4.6, while LocAgent's effective cost is \$1.65 per instance.
\end{observation}

\subsection{ReAct Agents for Iterative Reasoning}
While the ReAct agent employed in this study demonstrated clear improvements in bug localization compared to traditional RAG pipelines, its architecture introduces several inherent limitations. ReAct operates through a strictly sequential reasoning loop (\textit{Thought}~$\rightarrow$~\textit{Action}~$\rightarrow$~\textit{Observation}~$\rightarrow$~\textit{Answer/Repeat}), which constrains both scalability and responsiveness. Each reasoning step appends additional traces to the prompt, causing rapid context growth over multiple iterations. This iterative accumulation increases inference latency and memory overhead. For instance, the traditional RAG pipeline required approximately 10–25 seconds per issue on our local GPT-OSS-120B setup with Ollama, whereas the agentic RAG pipeline often required 15–40 seconds—particularly when the agent performed more reasoning steps before reaching a final decision. However, such issues were not noticed when using the Claude-4.6 model due to its fast inference.

To mitigate this overhead, our proposed pipeline incorporates two strategies. First, the agent operates on a limited candidate set (e.g., Top-15 retrieved files) to reduce unnecessary reasoning scope. Second, it is explicitly instructed to analyze detailed file contents (i.e., code skeletons) only when deemed necessary by its reasoning policy. Nonetheless, the ReAct architecture remains inherently sequential and lacks parallelism. \rev{Consequently, the agent's decision-making becomes tightly coupled with the linear order of its reasoning: an incorrect hypothesis formed in an early reasoning step---for example, prematurely concluding that a structurally similar file is the fault location---can bias subsequent file inspections and persist uncorrected through the final scoring output.} This strict sequentiality, while advantageous for interpretability and modular reasoning, imposes trade-offs in scenarios requiring long candidate pool exploration or frequent context switching.

\begin{observation}{obs:react-limitation}
The sequential nature of ReAct reasoning constrains scalability and responsiveness, causing early errors to propagate and increasing inference latency and memory usage.
\end{observation}

\subsection{Code Skeletons for Efficient Context Management}
To support effective agentic reasoning, we employed abstract code skeletons—structural representations containing only class and function signatures—instead of full source code. Prior work has shown that using code skeletons improves localization accuracy in LLM-based repair systems~\cite{xia2025agentless}. In our setting, however, this design choice is primarily motivated by the context sensitivity of ReAct agents, whose internal state grows cumulatively across reasoning turns. Specifically, each iteration in a ReAct agent’s cycle adds reasoning traces, action descriptions, and intermediate results to the conversation history. This cumulative expansion rapidly depletes the available context window and escalates computational cost, making ReAct agents expensive when many reasoning steps are required or the input is large.

During our experiments, we observed that limiting the agent’s view to the code skeleton substantially improved both reasoning efficiency and overall stability. When restricted to skeletons, the agent typically converged within five reasoning steps, as it was guided to reach a conclusion as soon as sufficient evidence was gathered. This behavior eliminates the need to sequentially rank all related files, allowing the agent to focus only on those it considers most relevant. In contrast, when provided with full source code files—often spanning thousands of tokens—the agent’s prompt grew rapidly, causing significant latency (sometimes exceeding ten minutes per action) and frequent failures due to context overflow.

Consequently, supplying structural representations rather than full implementations mitigates prompt saturation, reduces token redundancy, and allows deeper inspection of candidate files within a fixed computational budget. These observations indicate that for ReAct-style agentic bug localization systems, the use of code skeletons is not merely an optimization but a functional necessity for maintaining both efficiency and reliability.

\begin{observation}{obs:code-skeleton}
Using code skeletons instead of full source files is essential for ReAct agents, reducing context growth, improving reasoning efficiency, and ensuring stable localization outcomes.
\end{observation}

\subsection{Query Transformation for Retrieval in Agentic Pipeline}
\label{sec:query-trans-discussion}
Bug localization often suffers from a severe lexical gap between bug reports and source code \cite{niu2025deep}. While neural retrieval models (e.g., transformer-based encoders \cite{reimers2019sentence}) partially mitigate this issue through semantic embeddings, they still struggle to capture the structural and behavioral cues essential for accurate localization. To address this, \textsc{BLAgent} reformulates each bug report into two complementary queries. The \textit{structural} query extracts code-centric identifiers, while the \textit{behavioral} query abstracts the described symptoms—an approach shown to improve retrieval effectiveness \cite{samir2025improved} (Observation~\ref{obs:query-transform}, \ref{obs:query-transform-2}).

The candidates retrieved by these queries provide \textsc{BLAgent} with complementary, multi‑perspective evidence for file‑level localization. Each query generates its own ranked list of candidate files, allowing the agent to cross‑validate results based on structural reasoning (i.e., viewing the code structure). Prior studies confirm that exploiting structural identifiers like class and method names improves retrieval precision in bug localization \cite{saha2013improving}, and that bug reports that clearly articulate observed and expected behavior provide richer localization cues \cite{bettenburg2008makes}. However, while using multiple transformed queries is beneficial, expanding the candidate pool with too many queries or a very large candidate pool can result in increased latency, agent reasoning overhead, and token budget in the ReAct loop (Observations \ref{obs:react-limitation}, \ref{obs:code-skeleton}). Hence, to balance coverage and efficiency, we limit the transformations to two complementary queries, ensuring that we capture the most important cues without overloading the agent. We then merge only the unique top-ranked files, limiting the candidate pool to 15 files (see Section \ref{sec:agentic-reranking-design}), and the agent subsequently ranks the top-10 files among this pool.

\begin{observation}{obs:query-discussion}
Query transformations provide complementary retrieval signals and improve dense retrieval performance; however, limiting the number of transformations and the candidate pool is crucial for efficient and stable agentic reasoning.
\end{observation}

\subsection{Effect of Context Size on EAR Performance} 
\label{sec:context-size-effect}
\rev{To validate the design choices underlying Phase 2, we ablate two key hyperparameters: the number of candidate files passed to EAR (\#Files) and the number of retriever-highlighted chunks used to construct each file's pruned context (\#Chunks). Table~\ref{tab:chunk-ablation} reports Top-1, Top-3, and prompt token counts across four configurations. Increasing the number of files from 5 to 10 raises token usage by over 86\% without Top-1 improvement, suggesting that lower-ranked candidates carry little additional signal and only inflate the context size. Similarly, an increasing number of chunks from 5 to 10 adds 26--27\% token overhead with no noticeable accuracy gain, indicating that the retriever already surfaces the most relevant method bodies within the top-5 chunks and further expansion recovers no new evidence. Based on these insights, we find the Top-5 files and chunks optimal for Phase 2 reranking.}

\begin{table}[htbp]
\centering
\caption{Impact of number of files and chunks in Evidence-Anchored Reranking.}
\label{tab:chunk-ablation}
\small
\begin{tabular}{lccccc}
\toprule
\textbf{\#Files} & \textbf{\#Chunks} & \textbf{Top-1} & \textbf{Top-3} & \textbf{\#Prompt Tokens} \\
\midrule
5 & 5 & 0.786 & 0.923 & 15,973\\
5 & 10 & 0.780 & 0.923 & 20,264\\
10 & 5 & 0.773 & 0.910 & 29,804 \\
10 & 10 & 0.780 & 0.930 & 37,884\\
\bottomrule
\end{tabular}
\end{table}

\rev{Overall, these results highlight that accurate file localization with a RAG pipeline depends less on how much context is provided and more on how well that context is selected, a principle reflected across both reranking phases and the ablation results.}

\begin{observation}{obs:context-size-phase2}
Increasing context size raises cost without improving accuracy, indicating that effective localization in EAR relies on narrowing the search space rather than expanding context.
\end{observation}
\section{Recommendations}
Based on the observations \textbf{(O$\star$)} from our study of agentic RAG for bug localization and its impact on program repair, we provide targeted recommendations \textbf{(R$\star$)} organized by three overarching themes: (1) Retrieval Quality and Pipeline Architecture, (2) Hierarchical Localization and Fine-Grained Reasoning, and (3) Practical Utility. Each recommendation is grounded in specific findings and paired with actionable implementation strategies. Table~\ref{tab:observations-summary} summarizes the relationships among our findings, recommendations, and supporting observations.

\subsection{Theme 1. Retrieval Quality and Pipeline Architecture}
\recommendation{Dense retrieval recall fundamentally bounds a RAG or agentic RAG pipeline—when the correct file is absent from the candidates provided to the pipeline, no subsequent reasoning can compensate. Instead of scaling to larger models, practitioners may focus on improving embedding quality and retrieval mechanisms. Path-aware, code-structured chunking yields 20.4\% improvement over text-based chunking, and incorporating relative file paths adds another 16.9\% gain to retrieval accuracy in our experiments. Similar approaches may be explored for code-aware embeddings that preserve syntactic boundaries and augment chunks with hierarchical repository context (module/file paths) to enhance semantic recall at the file level.}

\recommendation{Retrieval pipelines should treat bug reports as multi-perspective queries rather than single text inputs. Refining queries has been found to be useful in RAG pipelines \cite{chan2024rq, lin2023decomposing}. Similarly, we also showed how transformations that disentangle structural (syntactic) and behavioral (semantic) aspects of a bug report allow the system to retrieve complementary code regions. This dual-channel formulation promotes balanced recall and precision, ensuring robust localization even when individual query views are incomplete. Similar approaches should be explored for better retrieval.}

\recommendation{While RAG pipelines improve over dense retrieval, agentic reranking allows for further improvement in file-level localization. \rev{Such pipelines can generalize to function-level with minimal adaptation, and cascading them with graph-guided function-level agents represents a promising direction toward fully hierarchical, cost-efficient localization for LLM-based patch generation.}}

\recommendation{We discussed in Section \ref{sec:llm-variation} that models with substantially different sizes can still yield competitive performance within an agentic pipeline when the right context is provided to the agent. This observation enables a practical approach for cost-conscious organizations: adopt tiered LLM utilization where medium-sized models (e.g., Qwen3-32B) handle localization through agentic reasoning, reserving larger or proprietary models (e.g., GPT-4) only for computationally demanding downstream tasks like patch generation, where semantic complexity may justify the additional expense.}

\begin{table}[h!]
\centering
\caption{Summary of Findings, Recommendations, and Supporting Observations}
\small
\begin{tabular}{p{7cm}|p{5.4cm}|p{2.1cm}}
\toprule
\textbf{Finding} & \textbf{Recommendation} & \textbf{Observation(s)} \\
\midrule
\multicolumn{3}{l}{\textbf{Theme 1: Retrieval Quality and Pipeline Architecture}} \\
\midrule
Dense retrieval recall bounds pipeline performance & R1: Prioritize retrieval enhancement & \ref{obs:localizeation-agentic-performance}, \ref{obs:chunking-performance}, \ref{obs:retrieval-determines} \\
\midrule
Query transformations provide complementary signals & R2: Multi-signal query transformation & \ref{obs:query-transform}, \ref{obs:query-transform-2} \\
\midrule
RAG improves robustness across Top-k, agentic RAG further improves across all Top-k ranks & R3: Leverage agentic reranking & \ref{obs:localization-performance}, \ref{obs:rag-improves}, \ref{obs:reranking-performance}, \ref{obs:query-transform}, \ref{obs:func-competitive}, \ref{obs:func-filelevel-foundation} \\
\midrule
Model scale provides minimal gains in agentic pipelines & R4: Smaller sized reasoning models should be evaluated & \ref{obs:llm-variation} \\
\midrule
\multicolumn{3}{l}{\textbf{Theme 2: Hierarchical Localization and Fine-Grained Reasoning}} \\
\midrule
Localization stages account for most of the failures & R5: Extend agentic reasoning to function/line level & \ref{obs:hierarchical-failure}, \ref{obs:loc-corrects-patch} \\
\midrule
Function-level parsing errors propagate failures & R6: Robust output parsing for functions & \ref{obs:parsing-failure} \\
\midrule
Line-level context insufficient; needs semantic reasoning & R7: Context-preserving line-level localization & \ref{obs:loc-corrects-patch} \\
\midrule
\multicolumn{3}{l}{\textbf{Theme 3: Practical Utility}} \\
\midrule
Semantic patch errors despite correct localization & R8: Semantic patch validation \& test integration & \ref{obs:repair-performance-when-test-integrated}, \ref{obs:patch-failure} \\
\midrule
Stochasticity causes unstable repair outcomes & R9: Stochasticity-aware evaluation metrics & \ref{obs:repair-diff} \\
\midrule
Modular integration improves APR effectiveness & R10: Modularize localization components & \ref{obs:repair-performance}, \ref{obs:repair-and-loc-relation} \\
\midrule
Agentic reasoning and large prompts may increase latency and instability & R11: Provide only necessary context and limit reasoning iterations & \ref{obs:react-limitation}, \ref{obs:code-skeleton}, \ref{obs:query-discussion} \\
\midrule
\rev{Agentic RAG is cost-efficient compared to unbounded graph-guided approaches} & \rev{R12: Use agentic RAG to narrow down search space and avoid unnecessary context expansion} & \ref{obs:func-competitive}, \ref{obs:cost-analysis}, \ref{obs:context-size-phase2} \\
\bottomrule
\end{tabular}
\label{tab:observations-summary}
\end{table}

\subsection{Theme 2. Hierarchical Localization and Fine-Grained Reasoning}

\recommendation{As shown in Figure~\ref{fig:repair-outcomes}b, function- and line-level localization comprise the majority of residual failures—even when file-level \textsc{BLAgent} succeeds. To address this critical bottleneck, we recommend extending reasoning methods to every hierarchical stage of localization. In particular, frameworks should (1) implement multi-step, structured reasoning to iteratively assess and rank function and line candidates, (2) deploy confidence-aware candidate expansion strategies—such as dynamically retrieving additional functions or statements when top candidates have low discriminatory signal, and (3) integrate program structure-aware validation checks that verify proposed localizations and edits maintain code correctness and semantic intent. Pursuing these reasoning-based context-rich approaches at all bug localization stages will close the gap between file-level success and end-to-end automated repair performance.}

\recommendation{APR frameworks should incorporate robust and tolerant output parsing strategies that leverage fuzzy matching or schema-driven validation to handle minor variations and formatting inconsistencies in LLM-generated outputs. Such techniques may reduce false negatives from strict parsing failures and enhance system reliability by enabling automatic corrections or fallback prompts. Empirical studies in LLM-based log parsing and structured data extraction highlight the effectiveness of these approaches in improving parsing robustness and downstream task performance \cite{ma2024llmparser}.}

\recommendation{Given that manual provision of correct line-level localization can improve automated repair success (see Figure \ref{fig:line-level-patch-fix}), APR systems may implement iterative refinement mechanisms for line-level localization. Specifically, these systems can dynamically generate, evaluate, and update candidate faulty lines through multi-pass reasoning or feedback loops, enabling correction of initial mislocalizations.}

\subsection{Theme 3. Practical Utility}

\recommendation{To improve APR success rates amidst semantic patch generation errors despite correct localization (Observation \ref{obs:patch-failure}), we strongly recommend tightly integrating testing and semantic validation within the patch selection pipeline. Specifically, regression and reproduction tests should be employed wherever available to filter generated patches before acceptance. When test oracles are unavailable, lightweight semantic validations—such as early syntax compilation checks, static analysis for invariant preservation, and heuristic differential testing against available reference implementations—should be incorporated.}

\recommendation{Given the inherent stochasticity of APR frameworks employing high-temperature decoding and sampling, as observed in our study with notably divergent repair sets across runs (Observation \ref{obs:repair-diff}), evaluations should incorporate multiple independent execution runs to capture variability. Rather than reporting only point estimates, practitioners are encouraged to aggregate results across multiple trials, analyze overlaps among uniquely repaired issues, and provide nuanced metrics that reflect both mean effectiveness and coverage diversity. }

\recommendation{The improvement in repair rates (Observation \ref{obs:repair-performance}) when integrating \textsc{BLAgent} into Agentless demonstrates that modular localization components can enhance existing APR systems. Framework developers may expose localization as a pluggable interface with standardized API contracts, allowing researchers and practitioners to quickly swap improved localization modules without modifying downstream components. This reduces integration friction and accelerates adoption of localization advances across the broader APR ecosystem.}

\recommendation{Agentic reasoning with ReAct architectures may face scalability challenges due to cumulative context growth and sequential reasoning constraints (Observations \ref{obs:react-limitation}, \ref{obs:code-skeleton}). To address this, if practitioners \textit{want to} adopt an agentic approach, they may (1) limit input context by providing lightweight structural abstractions such as code skeletons when possible, (2) implement adaptive iteration controls to encourage early convergence of reasoning, and (3) crucially incorporate explicit agent memory mechanisms \cite{zhang2025survey} that persist and selectively reuse past reasoning states, observations, and decisions. The integration of persistent memory allows the agent to flexibly update its internal state without reiterating the entire reasoning history, mitigating prompt saturation and error propagation.}

\recommendation{\rev{Unbounded graph-guided localization achieves strong accuracy but at substantial inference cost. Agentic RAG instead constrains reasoning to a compact, retrieval-filtered candidate set, avoiding unnecessary context expansion while achieving comparable accuracy. Practitioners may adopt agentic RAG as a low-cost upstream filter and reserve graph-guided traversal for low-confidence cases, enabling scalable hierarchical localization without sacrificing precision.}}
\section{Threats to Validity}

\noindent\textbf{Internal Validity.}
Internal validity concerns whether the observed improvements can be confidently attributed to the proposed agentic RAG framework rather than uncontrolled factors. To mitigate potential confounding effects, we maintained consistent experimental conditions across all pipelines, including identical retrieval databases, model configurations, and prompt templates. The ReAct agent’s reasoning cycles were constrained by a fixed step limit and identical prompting protocol to improve reproducibility. Nevertheless, since LLM inference remains stochastic under the selected temperature settings, intermediate reasoning behaviors may vary across runs. Future work could further reduce this variance by incorporating repeated trials and statistical aggregation of results.

\noindent\textbf{External Validity.}
Our evaluation relied primarily on the SWE-bench-Lite benchmark, which, while representative of Python-based GitHub projects and issues, may not fully reflect the diversity of industrial-scale software development. Projects with highly domain-specific APIs, sparse documentation, or unconventional code organization could produce different outcomes. Moreover, the agent’s effectiveness was assessed under specific model and retrieval settings (e.g., medium-sized LLMs with static retrieval backends). Extending the evaluation to other programming languages, larger repositories, and alternative retrieval or agent architectures would enhance the generalizability of our findings. \rev{Furthermore, our evaluation uses APR as a representative downstream task. While APR provides an objective and scalable setting, bug localization is also used in developer-centric tasks such as debugging and root-cause analysis, where evaluation typically requires user studies. Thus, our findings may not fully generalize to all usage scenarios. We leave broader evaluations as future work.}
\section{Related Work}
Our work bridges two active research areas: bug localization and automated program repair, with a particular emphasis on leveraging retrieval-augmented generation and agentic reasoning for repository-level bug localization. We review representative work in these areas and position our contributions.

\subsection{Bug Localization}
Bug localization techniques aim to identify buggy code locations to assist debugging. These approaches can be categorized into spectrum-based, information retrieval-based, and learning-based methods.

\noindent\textbf{Spectrum-Based Fault Localization (SBFL).}
SBFL techniques~\cite{wong2016survey,jones2002visualization,abreu2007accuracy} leverage program execution spectra—coverage information collected from passing and failing test cases—to compute a suspiciousness score for each program entity. Classic formulas such as Tarantula~\cite{jones2002visualization}, Ochiai~\cite{abreu2007accuracy}, and Dstar~\cite{wong2013dstar} quantify the likelihood that a statement or function is faulty. To improve on these traditional methods, Zhang et al.~\cite{zhang2019empirical} proposed PRFL, which incorporates a PageRank-based reweighting mechanism to better capture the relative contribution of individual tests, achieving notable accuracy gains over conventional SBFL formulations.

\noindent\textbf{Information Retrieval (IR)-Based Localization.}
IR-based approaches~\cite{zhou2012should,saha2013improving,lam2017bug,wang2014version,wang2015evaluating} model localization as a retrieval problem, ranking source files according to their textual similarity to bug reports. BugLocator~\cite{zhou2012should} pioneered this direction by introducing a revised Vector Space Model (rVSM) that integrates historical bug-fix information to enhance ranking. Subsequent work, such as Pathidea~\cite{chen2021pathidea}, augmented IR-based retrieval with execution path reconstruction from log data, demonstrating that runtime information can significantly complement textual similarity in identifying faulty files.

\noindent\textbf{Learning-Based Localization.}
More recently, deep learning methods have become increasingly prevalent~\cite{zhang2021study,li2019deepfl,meng2022improving}. Zhang et al.~\cite{zhang2021study} conducted a comprehensive empirical study comparing convolutional neural networks (CNNs), recurrent neural networks (RNNs), and multilayer perceptrons (MLPs) for real-world bug localization, finding CNN-based approaches most effective. Lam et al.~\cite{lam2017bug} integrated deep neural representations with VSM to mitigate lexical mismatch between bug reports and code artifacts.

The emergence of LLMs has further advanced this field by enabling models to jointly reason over natural language and source code~\cite{chang2025bridging,jiang2025cosil,asad2025leveraging,qin2024agentfl,wu2023large}. Recent studies have explored both \textit{agentic}~\cite{qin2024agentfl} and \textit{agent-less}~\cite{xia2025agentless,chang2025bridging,jimenez2024swebench} paradigms for repository-level bug localization, consistently outperforming traditional approaches. \rev{AutoFL~\cite{kang2024quantitative} is an early LLM-based approach that uses ChatGPT with tool calls to navigate Java projects and localize faults on Defects4J \cite{just2014defects4j}, given failing tests and their execution outcomes. AgentFL~\cite{qin2024agentfl} scales this setting to project-level context by coordinating multiple LLM-based agents for fault comprehension, code navigation, and confirmation, but assumes a test-driven Defects4J workflow. LocAgent~\cite{chen2025locagent} constructs a heterogeneous code graph over files, classes, and functions, and uses an LLM agent to navigate it iteratively via BM25-based entity search, graph 
traversal, and code retrieval, progressively narrowing the fault location from file to function-level on SWE-bench-Lite and LocBench. While effective, its multi-step graph traversal incurs substantially higher cost per instance than retrieval-based approaches according to our evaluation and literature \cite{jiang2025cosil}.}

To the best of our knowledge, no prior work has examined how RAG pipelines perform when coupled with an agentic reasoning process for \textit{file-level localization}. Moreover, existing studies have not explored whether transforming bug reports into retrieval-friendly queries can further improve ranking accuracy in agentic pipelines.

\subsection{Automated Program Repair}
Automated Program Repair (APR) techniques attempt to automatically generate patches to fix software bugs. While we do not propose a new APR technique or improve upon existing APR techniques, one effective way to measure the impact of an FL technique is to see how the downstream APR gets impacted by a localization method. Different approaches like template-based \cite{liu2019tbar}, heuristic-based \cite{le2016history}, or LLM-based \cite{xia2025agentless, wang2024openhands, zhang2024systematic} are utilized for automated program repair. Recent works have shown that learning-based frameworks adopting LLMs that generate multiple candidate patches for each bug outperforms all other traditional approaches, including neural machine translation (NMT) approaches \cite{xia2023automated}. 

LLM-based APR has evolved through four main paradigms--fine-tuning, prompting, procedural, and agentic \cite{yang2025survey}. Fine-tuning approaches~\cite{zhang2023coder,machavcek2025impact} adapt LLM weights using bug-fix data for task alignment, achieving strong performance but demanding high computational cost. Prompting methods~\cite{xia2023conversational,fan2023automated} rely on carefully designed prompts to elicit repairs from pre-trained models without retraining, offering flexibility but limited by prompt quality and context length. Procedural frameworks~\cite{xia2025agentless,jimenez2024swebench, zhao2025recode} decompose repair into explicit stages such as localization, patch generation, and validation, enabling reproducibility with moderate overhead. Agentic systems~\cite{yang2024swe,bouzenia2024repairagent,zhang2024autocoderover, wang2024openhands} instead delegate control to the LLM, allowing dynamic planning and reasoning for multi-file bugs at the cost of higher latency. Despite the impressive success of LLM-based APR frameworks, their accuracy in generating correct patches highly depends on correct FL \cite{li2025hybrid, chang2025bridging}. While existing LLM-based APR research has primarily focused on enhancing patch generation mechanisms and prompt engineering strategies, our work addresses the upstream bottleneck--\textit{file-level localization}, which may limit all these approaches.

\section{Conclusion \& Future Work}

In this paper, we presented \textsc{BLAgent}, a novel agentic retrieval-augmented generation (RAG) framework for file-level bug localization that combines structure-aware retrieval, dual-perspective query transformation, and bounded agentic reasoning. By explicitly modeling both structural and behavioral signals from bug reports and grounding reasoning over retrieval-filtered code contexts, \textsc{BLAgent} achieves state-of-the-art localization performance on SWE-bench-Lite while maintaining substantially lower cost compared to existing graph-based or heavily instrumented approaches. Our findings further demonstrate that improvements at the file level translate directly into meaningful gains in downstream automated program repair. Furthermore, the proposed framework generalizes across localization granularities, as our results show that the same agentic RAG design naturally extends to finer-grained tasks such as function-level localization, highlighting its flexibility as a unified reasoning framework over code.

\rev{Looking forward, a key direction is to evolve BLAgent into a unified, multi-granularity localization framework that seamlessly integrates file-, function-, and line-level reasoning within a single pipeline. While the current design naturally extends across these levels, further optimization is needed to enable precise fine-grained localization without compromising efficiency. In particular, we envision a hybrid retrieval strategy where dense retrieval is complemented with lightweight graph-based exploration in cases where semantic signals are insufficient, enabling more robust candidate discovery in complex dependency structures. Coupled with tighter integration between retrieval and reasoning, this would allow the agent to progressively refine localization from coarse to fine granularity, moving toward an end-to-end localization system that directly supports debugging and automated repair.}

\section{Data Availability}
The replication package for this study, including the implementation, experimental scripts, and supporting materials, is publicly available at: \url{https://github.com/afifaniks/BLAgent}.
\bibliographystyle{ACM-Reference-Format}
\bibliography{references}

@article{xiao2019improving,
  title={Improving bug localization with word embedding and enhanced convolutional neural networks},
  author={Xiao, Yan and Keung, Jacky and Bennin, Kwabena E and Mi, Qing},
  journal={Information and Software Technology},
  volume={105},
  pages={17--29},
  year={2019},
  publisher={Elsevier}
}

@article{shao2024enhancing,
  title={Enhancing IR-based Fault Localization using Large Language Models},
  author={Shao, Shuai and Yu, Tingting},
  journal={arXiv preprint arXiv:2412.03754},
  year={2024}
}

@article{jiang2025cosil,
  title={CoSIL: Software Issue Localization via LLM-Driven Code Repository Graph Searching},
  author={Jiang, Zhonghao and Ren, Xiaoxue and Yan, Meng and Jiang, Wei and Li, Yong and Liu, Zhongxin},
  journal={arXiv preprint arXiv:2503.22424},
  year={2025}
}

@article{chang2025bridging,
  title={Bridging bug localization and issue fixing: A hierarchical localization framework leveraging large language models},
  author={Chang, Jianming and Zhou, Xin and Lulu, Lulu and Lo, David and Li, Bixin},
  journal={IEEE Transactions on Software Engineering},
  year={2026},
  publisher={IEEE}
}

@article{xia2025agentless,
author = {Xia, Chunqiu Steven and Deng, Yinlin and Dunn, Soren and Zhang, Lingming},
title = {Demystifying LLM-Based Software Engineering Agents},
year = {2025},
issue_date = {July 2025},
publisher = {Association for Computing Machinery},
address = {New York, NY, USA},
volume = {2},
number = {FSE},
url = {https://doi.org/10.1145/3715754},
doi = {10.1145/3715754},
journal = {Proc. ACM Softw. Eng.},
month = jun,
articleno = {FSE037},
numpages = {24}
}

@article{chan2024rq,
  title={Rq-rag: Learning to refine queries for retrieval augmented generation},
  author={Chan, Chi-Min and Xu, Chunpu and Yuan, Ruibin and Luo, Hongyin and Xue, Wei and Guo, Yike and Fu, Jie},
  journal={arXiv preprint arXiv:2404.00610},
  year={2024}
}

@article{li2024dmqr,
  title={Dmqr-rag: Diverse multi-query rewriting for rag},
  author={Li, Zhicong and Wang, Jiahao and Jiang, Zhishu and Mao, Hangyu and Chen, Zhongxia and Du, Jiazhen and Zhang, Yuanxing and Zhang, Fuzheng and Zhang, Di and Liu, Yong},
  journal={arXiv preprint arXiv:2411.13154},
  year={2024}
}

@inproceedings{ma2023query,
  title={Query rewriting in retrieval-augmented large language models},
  author={Ma, Xinbei and Gong, Yeyun and He, Pengcheng and Duan, Nan and others},
  booktitle={The 2023 Conference on Empirical Methods in Natural Language Processing},
  year={2023}
}

@inproceedings{yao2022react,
  title={React: Synergizing reasoning and acting in language models},
  author={Yao, Shunyu and Zhao, Jeffrey and Yu, Dian and Du, Nan and Shafran, Izhak and Narasimhan, Karthik R and Cao, Yuan},
  booktitle={The eleventh international conference on learning representations},
  year={2022}
}

@inproceedings{
jimenez2024swebench,
title={{SWE}-bench: Can Language Models Resolve Real-world Github Issues?},
author={Carlos E Jimenez and John Yang and Alexander Wettig and Shunyu Yao and Kexin Pei and Ofir Press and Karthik R Narasimhan},
booktitle={The Twelfth International Conference on Learning Representations},
year={2024},
url={https://openreview.net/forum?id=VTF8yNQM66}
}

@article{wang2024openhands,
  title={Openhands: An open platform for ai software developers as generalist agents},
  author={Wang, Xingyao and Li, Boxuan and Song, Yufan and Xu, Frank F and Tang, Xiangru and Zhuge, Mingchen and Pan, Jiayi and Song, Yueqi and Li, Bowen and Singh, Jaskirat and others},
  journal={arXiv preprint arXiv:2407.16741},
  year={2024}
}

@article{yang2024swe,
  title={Swe-agent: Agent-computer interfaces enable automated software engineering},
  author={Yang, John and Jimenez, Carlos E and Wettig, Alexander and Lieret, Kilian and Yao, Shunyu and Narasimhan, Karthik and Press, Ofir},
  journal={Advances in Neural Information Processing Systems},
  volume={37},
  pages={50528--50652},
  year={2024}
}

@article{liu2023lost,
  title={Lost in the middle: How language models use long contexts},
  author={Liu, Nelson F and Lin, Kevin and Hewitt, John and Paranjape, Ashwin and Bevilacqua, Michele and Petroni, Fabio and Liang, Percy},
  journal={arXiv preprint arXiv:2307.03172},
  year={2023}
}

@article{lu2021fantastically,
  title={Fantastically ordered prompts and where to find them: Overcoming few-shot prompt order sensitivity},
  author={Lu, Yao and Bartolo, Max and Moore, Alastair and Riedel, Sebastian and Stenetorp, Pontus},
  journal={arXiv preprint arXiv:2104.08786},
  year={2021}
}

@inproceedings{joshi2023repair,
  title={Repair is nearly generation: Multilingual program repair with llms},
  author={Joshi, Harshit and Sanchez, Jos{\'e} Cambronero and Gulwani, Sumit and Le, Vu and Verbruggen, Gust and Radi{\v{c}}ek, Ivan},
  booktitle={Proceedings of the AAAI Conference on Artificial Intelligence},
  volume={37},
  number={4},
  pages={5131--5140},
  year={2023}
}

@article{zhang2025cast,
  title={cAST: Enhancing Code Retrieval-Augmented Generation with Structural Chunking via Abstract Syntax Tree},
  author={Zhang, Yilin and Zhao, Xinran and Wang, Zora Zhiruo and Yang, Chenyang and Wei, Jiayi and Wu, Tongshuang},
  journal={arXiv preprint arXiv:2506.15655},
  year={2025}
}

@article{zhao2025recode,
  title={ReCode: Improving LLM-based Code Repair with Fine-Grained Retrieval-Augmented Generation},
  author={Zhao, Yicong and Chen, Shisong and Zhang, Jiacheng and Li, Zhixu},
  journal={arXiv preprint arXiv:2509.02330},
  year={2025}
}

@article{tao2025retrieval,
  title={Retrieval-Augmented Code Generation: A Survey with Focus on Repository-Level Approaches},
  author={Tao, Yicheng and Qin, Yao and Liu, Yepang},
  journal={arXiv preprint arXiv:2510.04905},
  year={2025}
}

@article{wong2016survey,
  title={A survey on software fault localization},
  author={Wong, W Eric and Gao, Ruizhi and Li, Yihao and Abreu, Rui and Wotawa, Franz},
  journal={IEEE Transactions on Software Engineering},
  volume={42},
  number={8},
  pages={707--740},
  year={2016},
  publisher={IEEE}
}

@inproceedings{abreu2007accuracy,
  title={On the accuracy of spectrum-based fault localization},
  author={Abreu, Rui and Zoeteweij, Peter and Van Gemund, Arjan JC},
  booktitle={Testing: Academic and industrial conference practice and research techniques-MUTATION (TAICPART-MUTATION 2007)},
  pages={89--98},
  year={2007},
  organization={IEEE}
}

@inproceedings{jones2002visualization,
  title={Visualization of test information to assist fault localization},
  author={Jones, James A and Harrold, Mary Jean and Stasko, John},
  booktitle={Proceedings of the 24th international conference on Software engineering},
  pages={467--477},
  year={2002}
}

@article{wong2013dstar,
  title={The DStar method for effective software fault localization},
  author={Wong, W Eric and Debroy, Vidroha and Gao, Ruizhi and Li, Yihao},
  journal={IEEE Transactions on Reliability},
  volume={63},
  number={1},
  pages={290--308},
  year={2013},
  publisher={IEEE}
}

@article{zhang2019empirical,
  title={An empirical study of boosting spectrum-based fault localization via pagerank},
  author={Zhang, Mengshi and Li, Yaoxian and Li, Xia and Chen, Lingchao and Zhang, Yuqun and Zhang, Lingming and Khurshid, Sarfraz},
  journal={IEEE Transactions on Software Engineering},
  volume={47},
  number={6},
  pages={1089--1113},
  year={2019},
  publisher={IEEE}
}

@inproceedings{zhou2012should,
  title={Where should the bugs be fixed? more accurate information retrieval-based bug localization based on bug reports},
  author={Zhou, Jian and Zhang, Hongyu and Lo, David},
  booktitle={2012 34th International conference on software engineering (ICSE)},
  pages={14--24},
  year={2012},
  organization={IEEE}
}

@inproceedings{saha2013improving,
  title={Improving bug localization using structured information retrieval},
  author={Saha, Ripon K and Lease, Matthew and Khurshid, Sarfraz and Perry, Dewayne E},
  booktitle={2013 28th IEEE/ACM International Conference on Automated Software Engineering (ASE)},
  pages={345--355},
  year={2013},
  organization={IEEE}
}

@inproceedings{lam2017bug,
  title={Bug localization with combination of deep learning and information retrieval},
  author={Lam, An Ngoc and Nguyen, Anh Tuan and Nguyen, Hoan Anh and Nguyen, Tien N},
  booktitle={2017 IEEE/ACM 25th International Conference on Program Comprehension (ICPC)},
  pages={218--229},
  year={2017},
  organization={IEEE}
}

@inproceedings{wang2014version,
  title={Version history, similar report, and structure: Putting them together for improved bug localization},
  author={Wang, Shaowei and Lo, David},
  booktitle={Proceedings of the 22nd international conference on program comprehension},
  pages={53--63},
  year={2014}
}

@inproceedings{wang2015evaluating,
  title={Evaluating the usefulness of ir-based fault localization techniques},
  author={Wang, Qianqian and Parnin, Chris and Orso, Alessandro},
  booktitle={Proceedings of the 2015 international symposium on software testing and analysis},
  pages={1--11},
  year={2015}
}

@article{chen2021pathidea,
  title={Pathidea: Improving information retrieval-based bug localization by re-constructing execution paths using logs},
  author={Chen, An Ran and Chen, Tse-Hsun and Wang, Shaowei},
  journal={IEEE Transactions on Software Engineering},
  volume={48},
  number={8},
  pages={2905--2919},
  year={2021},
  publisher={IEEE}
}

@article{zhang2021study,
  title={A study of effectiveness of deep learning in locating real faults},
  author={Zhang, Zhuo and Lei, Yan and Mao, Xiaoguang and Yan, Meng and Xu, Ling and Zhang, Xiaohong},
  journal={Information and Software Technology},
  volume={131},
  pages={106486},
  year={2021},
  publisher={Elsevier}
}

@inproceedings{li2019deepfl,
  title={Deepfl: Integrating multiple fault diagnosis dimensions for deep fault localization},
  author={Li, Xia and Li, Wei and Zhang, Yuqun and Zhang, Lingming},
  booktitle={Proceedings of the 28th ACM SIGSOFT international symposium on software testing and analysis},
  pages={169--180},
  year={2019}
}

@inproceedings{meng2022improving,
  title={Improving fault localization and program repair with deep semantic features and transferred knowledge},
  author={Meng, Xiangxin and Wang, Xu and Zhang, Hongyu and Sun, Hailong and Liu, Xudong},
  booktitle={Proceedings of the 44th International Conference on Software Engineering},
  pages={1169--1180},
  year={2022}
}

@article{asad2025leveraging,
  title={Leveraging large language model for information retrieval-based bug localization},
  author={Asad, Moumita and Yasir, Rafed Muhammad and Geramirad, Armin and Malek, Sam},
  journal={arXiv preprint arXiv:2508.00253},
  year={2025}
}

@article{qin2024agentfl,
  title={Agentfl: Scaling llm-based fault localization to project-level context},
  author={Qin, Yihao and Wang, Shangwen and Lou, Yiling and Dong, Jinhao and Wang, Kaixin and Li, Xiaoling and Mao, Xiaoguang},
  journal={arXiv preprint arXiv:2403.16362},
  year={2024}
}

@article{wu2023large,
  title={Large language models in fault localisation},
  author={Wu, Yonghao and Li, Zheng and Zhang, Jie M and Papadakis, Mike and Harman, Mark and Liu, Yong},
  journal={arXiv preprint arXiv:2308.15276},
  year={2023}
}

@inproceedings{xia2023automated,
  title={Automated program repair in the era of large pre-trained language models},
  author={Xia, Chunqiu Steven and Wei, Yuxiang and Zhang, Lingming},
  booktitle={2023 IEEE/ACM 45th International Conference on Software Engineering (ICSE)},
  pages={1482--1494},
  year={2023},
  organization={IEEE}
}

@article{yang2025survey,
  title={A Survey of LLM-based Automated Program Repair: Taxonomies, Design Paradigms, and Applications},
  author={Yang, Boyang and Cai, Zijian and Liu, Fengling and Le, Bach and Zhang, Lingming and Bissyand{\'e}, Tegawend{\'e} F and Liu, Yang and Tian, Haoye},
  journal={arXiv preprint arXiv:2506.23749},
  year={2025}
}

@inproceedings{zhang2023coder,
  title={Coder reviewer reranking for code generation},
  author={Zhang, Tianyi and Yu, Tao and Hashimoto, Tatsunori and Lewis, Mike and Yih, Wen-tau and Fried, Daniel and Wang, Sida},
  booktitle={International Conference on Machine Learning},
  pages={41832--41846},
  year={2023},
  organization={PMLR}
}

@article{machavcek2025impact,
  title={The Impact of Fine-tuning Large Language Models on Automated Program Repair},
  author={Mach{\'a}{\v{c}}ek, Roman and Grishina, Anastasiia and Hort, Max and Moonen, Leon},
  journal={arXiv preprint arXiv:2507.19909},
  year={2025}
}

@article{xia2023conversational,
  title={Conversational automated program repair},
  author={Xia, Chunqiu Steven and Zhang, Lingming},
  journal={arXiv preprint arXiv:2301.13246},
  year={2023}
}

@inproceedings{fan2023automated,
  title={Automated repair of programs from large language models},
  author={Fan, Zhiyu and Gao, Xiang and Mirchev, Martin and Roychoudhury, Abhik and Tan, Shin Hwei},
  booktitle={2023 IEEE/ACM 45th International Conference on Software Engineering (ICSE)},
  pages={1469--1481},
  year={2023},
  organization={IEEE}
}

@inproceedings{zhang2024autocoderover,
  title={Autocoderover: Autonomous program improvement},
  author={Zhang, Yuntong and Ruan, Haifeng and Fan, Zhiyu and Roychoudhury, Abhik},
  booktitle={Proceedings of the 33rd ACM SIGSOFT International Symposium on Software Testing and Analysis},
  pages={1592--1604},
  year={2024}
}

@article{bouzenia2024repairagent,
  title={Repairagent: An autonomous, llm-based agent for program repair},
  author={Bouzenia, Islem and Devanbu, Premkumar and Pradel, Michael},
  journal={arXiv preprint arXiv:2403.17134},
  year={2024}
}

@article{li2025hybrid,
  title={Hybrid automated program repair by combining large language models and program analysis},
  author={Li, Fengjie and Jiang, Jiajun and Sun, Jiajun and Zhang, Hongyu},
  journal={ACM Transactions on Software Engineering and Methodology},
  volume={34},
  number={7},
  pages={1--28},
  year={2025},
  publisher={ACM New York, NY}
}

@article{wong2023software,
  title={Software fault localization: An overview of research, techniques, and tools},
  author={Wong, W Eric and Gao, Ruizhi and Li, Yihao and Abreu, Rui and Wotawa, Franz and Li, Dongcheng},
  journal={Handbook of Software Fault Localization: Foundations and Advances},
  pages={1--117},
  year={2023},
  publisher={Wiley Online Library}
}

@article{lewis2020retrieval,
  title={Retrieval-augmented generation for knowledge-intensive nlp tasks},
  author={Lewis, Patrick and Perez, Ethan and Piktus, Aleksandra and Petroni, Fabio and Karpukhin, Vladimir and Goyal, Naman and K{\"u}ttler, Heinrich and Lewis, Mike and Yih, Wen-tau and Rockt{\"a}schel, Tim and others},
  journal={Advances in neural information processing systems},
  volume={33},
  pages={9459--9474},
  year={2020}
}

@article{huang2024understanding,
  title={Understanding the planning of LLM agents: A survey},
  author={Huang, Xu and Liu, Weiwen and Chen, Xiaolong and Wang, Xingmei and Wang, Hao and Lian, Defu and Wang, Yasheng and Tang, Ruiming and Chen, Enhong},
  journal={arXiv preprint arXiv:2402.02716},
  year={2024}
}

@article{guo2024large,
  title={Large language model based multi-agents: A survey of progress and challenges},
  author={Guo, Taicheng and Chen, Xiuying and Wang, Yaqi and Chang, Ruidi and Pei, Shichao and Chawla, Nitesh V and Wiest, Olaf and Zhang, Xiangliang},
  journal={arXiv preprint arXiv:2402.01680},
  year={2024}
}

@article{parvez2021retrieval,
  title={Retrieval augmented code generation and summarization},
  author={Parvez, Md Rizwan and Ahmad, Wasi Uddin and Chakraborty, Saikat and Ray, Baishakhi and Chang, Kai-Wei},
  journal={arXiv preprint arXiv:2108.11601},
  year={2021}
}

@inproceedings{sawarkar2024blended,
  title={Blended rag: Improving rag (retriever-augmented generation) accuracy with semantic search and hybrid query-based retrievers},
  author={Sawarkar, Kunal and Mangal, Abhilasha and Solanki, Shivam Raj},
  booktitle={2024 IEEE 7th international conference on multimedia information processing and retrieval (MIPR)},
  pages={155--161},
  year={2024},
  organization={IEEE}
}

@article{malkov2018efficient,
  title={Efficient and robust approximate nearest neighbor search using hierarchical navigable small world graphs},
  author={Malkov, Yu A and Yashunin, Dmitry A},
  journal={IEEE transactions on pattern analysis and machine intelligence},
  volume={42},
  number={4},
  pages={824--836},
  year={2018},
  publisher={IEEE}
}

@article{shazeer2017outrageously,
  title={Outrageously large neural networks: The sparsely-gated mixture-of-experts layer},
  author={Shazeer, Noam and Mirhoseini, Azalia and Maziarz, Krzysztof and Davis, Andy and Le, Quoc and Hinton, Geoffrey and Dean, Jeff},
  journal={arXiv preprint arXiv:1701.06538},
  year={2017}
}

@article{huang2025survey,
  title={A survey on hallucination in large language models: Principles, taxonomy, challenges, and open questions},
  author={Huang, Lei and Yu, Weijiang and Ma, Weitao and Zhong, Weihong and Feng, Zhangyin and Wang, Haotian and Chen, Qianglong and Peng, Weihua and Feng, Xiaocheng and Qin, Bing and others},
  journal={ACM Transactions on Information Systems},
  volume={43},
  number={2},
  pages={1--55},
  year={2025},
  publisher={ACM New York, NY}
}

@article{lin2023decomposing,
  title={Decomposing complex queries for tip-of-the-tongue retrieval},
  author={Lin, Kevin and Lo, Kyle and Gonzalez, Joseph E and Klein, Dan},
  journal={arXiv preprint arXiv:2305.15053},
  year={2023}
}

@inproceedings{ma2024llmparser,
  title={Llmparser: An exploratory study on using large language models for log parsing},
  author={Ma, Zeyang and Chen, An Ran and Kim, Dong Jae and Chen, Tse-Hsun and Wang, Shaowei},
  booktitle={Proceedings of the IEEE/ACM 46th International Conference on Software Engineering},
  pages={1--13},
  year={2024}
}

@article{zhang2025survey,
  title={A survey on the memory mechanism of large language model-based agents},
  author={Zhang, Zeyu and Dai, Quanyu and Bo, Xiaohe and Ma, Chen and Li, Rui and Chen, Xu and Zhu, Jieming and Dong, Zhenhua and Wen, Ji-Rong},
  journal={ACM Transactions on Information Systems},
  volume={43},
  number={6},
  pages={1--47},
  year={2025},
  publisher={ACM New York, NY}
}

@article{niu2025deep,
  title={When Deep Learning Meets Information Retrieval-based Bug Localization: A Survey},
  author={Niu, Feifei and Li, Chuanyi and Liu, Kui and Xia, Xin and Lo, David},
  journal={ACM Computing Surveys},
  volume={57},
  number={11},
  pages={1--41},
  year={2025},
  publisher={ACM New York, NY}
}

@article{reimers2019sentence,
  title={Sentence-bert: Sentence embeddings using siamese bert-networks},
  author={Reimers, Nils and Gurevych, Iryna},
  journal={arXiv preprint arXiv:1908.10084},
  year={2019}
}

@article{samir2025improved,
  title={Improved IR-based Bug Localization with Intelligent Relevance Feedback},
  author={Samir, Asif Mohammed and Rahman, Mohammad Masudur},
  journal={arXiv preprint arXiv:2501.10542},
  year={2025}
}

@inproceedings{bettenburg2008makes,
  title={What makes a good bug report?},
  author={Bettenburg, Nicolas and Just, Sascha and Schr{\"o}ter, Adrian and Weiss, Cathrin and Premraj, Rahul and Zimmermann, Thomas},
  booktitle={Proceedings of the 16th ACM SIGSOFT International Symposium on Foundations of software engineering},
  pages={308--318},
  year={2008}
}

@inproceedings{liu2019tbar,
  title={TBar: Revisiting template-based automated program repair},
  author={Liu, Kui and Koyuncu, Anil and Kim, Dongsun and Bissyand{\'e}, Tegawend{\'e} F},
  booktitle={Proceedings of the 28th ACM SIGSOFT international symposium on software testing and analysis},
  pages={31--42},
  year={2019}
}

@inproceedings{le2016history,
  title={History driven program repair},
  author={Le, Xuan Bach D and Lo, David and Le Goues, Claire},
  booktitle={2016 IEEE 23rd international conference on software analysis, evolution, and reengineering (SANER)},
  volume={1},
  pages={213--224},
  year={2016},
  organization={IEEE}
}

@article{zhang2024systematic,
  title={A systematic literature review on large language models for automated program repair},
  author={Zhang, Quanjun and Fang, Chunrong and Xie, Yang and Ma, YuXiang and Sun, Weisong and Yang, Yun and Chen, Zhenyu},
  journal={arXiv preprint arXiv:2405.01466},
  year={2024}
}

@inproceedings{chen2025locagent,
  title={Locagent: Graph-guided llm agents for code localization},
  author={Chen, Zhaoling and Tang, Robert and Deng, Gangda and Wu, Fang and Wu, Jialong and Jiang, Zhiwei and Prasanna, Viktor and Cohan, Arman and Wang, Xingyao},
  booktitle={Proceedings of the 63rd Annual Meeting of the Association for Computational Linguistics (Volume 1: Long Papers)},
  pages={8697--8727},
  year={2025}
}

@inproceedings{qu2025semantic,
  title={Is semantic chunking worth the computational cost?},
  author={Qu, Renyi and Tu, Ruixuan and Bao, Forrest},
  booktitle={Findings of the Association for Computational Linguistics: NAACL 2025},
  pages={2155--2177},
  year={2025}
}

@article{kang2024quantitative,
  title={A quantitative and qualitative evaluation of LLM-based explainable fault localization},
  author={Kang, Sungmin and An, Gabin and Yoo, Shin},
  journal={Proceedings of the ACM on Software Engineering},
  volume={1},
  number={FSE},
  pages={1424--1446},
  year={2024},
  publisher={ACM New York, NY, USA}
}

@inproceedings{just2014defects4j,
author = {Just, Ren\'{e} and Jalali, Darioush and Ernst, Michael D.},
title = {Defects4J: a database of existing faults to enable controlled testing studies for Java programs},
year = {2014},
isbn = {9781450326452},
publisher = {Association for Computing Machinery},
address = {New York, NY, USA},
url = {https://doi.org/10.1145/2610384.2628055},
doi = {10.1145/2610384.2628055},
booktitle = {Proceedings of the 2014 International Symposium on Software Testing and Analysis},
pages = {437–440},
numpages = {4},
location = {San Jose, CA, USA},
series = {ISSTA 2014}
}

@inproceedings{ma2025alibaba,
  title={Alibaba lingmaagent: Improving automated issue resolution via comprehensive repository exploration},
  author={Ma, Yingwei and Yang, Qingping and Cao, Rongyu and Li, Binhua and Huang, Fei and Li, Yongbin},
  booktitle={Proceedings of the 33rd ACM International Conference on the Foundations of Software Engineering},
  pages={238--249},
  year={2025}
}

@inproceedings{bohme2017bug,
  title={Where is the bug and how is it fixed? an experiment with practitioners},
  author={B{\"o}hme, Marcel and Soremekun, Ezekiel O and Chattopadhyay, Sudipta and Ugherughe, Emamurho and Zeller, Andreas},
  booktitle={Proceedings of the 2017 11th joint meeting on foundations of software engineering},
  pages={117--128},
  year={2017}
}

@misc{stripe2018developer,
  title        = {The Developer Coefficient: Software Engineering Efficiency 
                  and Its {{\$}}3 Trillion Impact on Global {GDP}},
  author       = {{Stripe}},
  year         = {2018},
  month        = sep,
  howpublished = {\url{https://stripe.com/files/reports/the-developer-coefficient.pdf}},
  note = {Accessed: 2026-03-24}
}

\end{document}